\documentclass[aps,pre,graphicx,reprint,amssymb,showpacs,superscriptaddress,floatfix,nobalancelastpage]{revtex4-1}
\usepackage[utf8]{inputenc}
\usepackage[T1]{fontenc}
\usepackage[english]{babel}
\usepackage{amsmath}
\usepackage{amsfonts}
\usepackage{amssymb,textcomp}
\usepackage[xcdraw]{xcolor}
\usepackage{colortbl}
\usepackage{tikz}
\usepackage{bm}
\usepackage{framed}
\usepackage{graphicx, wrapfig}
\usepackage{array,multirow}
\usepackage{lmodern}					
\usepackage{comment}
\usepackage{natbib}
\usepackage[margin=2cm]{geometry}
\usepackage{hyperref}

\usepackage{stackengine}
\usepackage{enumitem}
\usepackage{mathtools}

\newcommand\beq{\begin{equation}}      
\newcommand\beqnn{\begin{eqnarray*}}   
\newcommand\beqa{\begin{eqnarray}}     
\newcommand\beqann{\begin{eqnarray*}}  

\newcommand\eeq{\end{equation}}        
\newcommand\eeqnn{\end{eqnarray*}}     
\newcommand\eeqa{\end{eqnarray}}       
\newcommand\eeqann{\end{eqnarray*}}    

\def\nl {\nonumber \\}

\newcommand\bi{\begin{itemize}}
\newcommand\ei{\end{itemize}}

\def\nl {\nonumber \\}

\newcommand{\eref}[1]{(\ref{#1})}

\newcommand{\fref}[1]{Figure~\ref{#1}}

\newcommand{\Eref}[1]{Eq.~(\ref{#1})}
\newcommand{\Sref}[1]{Sec.~\ref{#1}}
\newcommand{\Fref}[1]{Figure~\ref{#1}}
\newcommand{\al}[1]{\begin{align} #1 \end{align}}
\def\l0{L}

\def\xt{\bm x,t}

\renewcommand{\vec}[1]{\bm{#1}}

\newcommand*\colvec[3][]{
    \begin{pmatrix}\ifx\relax#1\relax\else#1\\\fi#2\\#3\end{pmatrix}
}

\newcommand{\com}[1]{\textcolor{blue}{#1}}

\def\bulk{^{(b)}}
\def\onew{^{(1w)}}
\def\twow{^{(2w)}}
\def\cyc{_C} 
\newcommand{\cmmnt}[1]{\ignorespaces}
\def\xt{\tilde x}

\def\zt{\tilde z}
\def\vt{\tilde v}
\def\mut{\tilde \mu}
\def\ts{t^*}
\def\taus{\tau^*}
\def\tsb{(t^*)}
\def\lt{\tilde\ell}
\def\alphat{\tilde\alpha}
\def\omegat{\tilde\omega}

\begin{document}

\title{Activated diffusiophoresis}
\author{Christian M. Rohwer}
\email[]{crohwer@is.mpg.de}
\affiliation{Max Planck Institute for Intelligent Systems, Heisenbergstr. 3, 70569 Stuttgart, Germany}
\affiliation{4th Institute for Theoretical Physics, University of Stuttgart, Pfaffenwaldring 57, 70569 Stuttgart, Germany}

\author{Mehran Kardar}
\affiliation{Department of Physics, Massachusetts Institute of Technology, Cambridge, Massachusetts 02139, USA}

\author{Matthias Kr\"uger}
\affiliation{Institute for Theoretical Physics, University of G\"ottingen, D-37077, G\"ottingen, Germany}

\date{\today}

\begin{abstract}
	Perturbations of fluid media can give rise to non-equilibrium dynamics, which may in turn cause motion of immersed inclusions. We consider perturbations (``activations'') that are local in space and time, of a fluid density which is conserved, and  study the resulting diffusiophoretic phenomena that emerge at a large distance. Specifically, we consider cases where the perturbations propagate diffusively,  providing examples from passive and active matter for which this is expected to be the case. Activations can, for instance, be realized by sudden and local changes in interaction potentials of the medium, or by local changes of its activity. Various analytical results are provided for the case of confinement by two parallel walls. We investigate the possibility of extracting work from inclusions which are moving through the activated fluid. Further, we show that a time-dependent density profile, created via suitable activation protocols, allows for  conveyance of inclusions along controlled and stable trajectories. In contrast, in states with a steady density, inclusions cannot be held at stable positions, reminiscent of Earnshaw's theorem of electrostatics.  We expect these findings to be applicable in a range of experimental systems.

\end{abstract}

\maketitle

\section{Introduction}
\label{sec:intro}

In colloid physics, phoresis refers to transport or motion of colloidal particles, typically induced by non-equilibrium conditions in the medium~\cite{anderson1989colloid}. Paradigmatic examples include diffusio-~\cite{derjaguin1947kinetic,anderson1986transport,velegol2016origins}, thermo-~\cite{soret1879,mcnab1973thermophoresis,somasundaran2006encyclopedia,piazza2008thermophoresis} and electrophoresis~\cite{smoluchowski1903contribution,russel1991colloidal,delgado2001interfacial}, which are due to gradients in density, temperature, or due to applied electric fields, respectively. 
Phoretic mechanisms based on external driving such as electric fields~\cite{loget2011electric} or temperature gradients~\cite{wurger2010ThermalNoneqTransp} have been utilized extensively to manipulate particles at small scales via optical traps~\cite{neuman2004optical} or optical conveyors~\cite{weinert2009OpticalConveyor}, and to trap individual molecules~\cite{cichos2015SingleMolTrap} and even proteins~\cite{franzl2019thermophoretic} through feedback control of externally imposed temperature fields. 
On the other hand, self-phoresis, which also garnered a lot of recent interest, involves processes where the colloid itself induces motion by generating fields or gradients within its fluid surroundings~\cite{golestanian2007DesigningSelfPhoretic}, for instance, through asymmetrical chemical reactions~\cite{golestanian2005SelfDiffusio,popescu2016},  locally induced temperature gradients~\cite{jiang2010SelfThermoPhor} or local demixing of a multi-component solvent~\cite{buttinoni2012active} at its surface. A wealth of chemically powered microswimmers have also been realized experimentally~\cite{ebbens2010ReviewPropulsion, moran2017phoretic}. In general, understanding phoretic phenomena remains a major challenge~\cite{moran2017phoretic}, due to the many microscopic mechanisms that can give rise to it~\cite{ebbens2010ReviewPropulsion,lauga2009HydrodynSwim}. 

In this manuscript we investigate a mechanism of phoresis which results from externally induced, spatially local, and temporally abrupt perturbations of a fluid
with a single conserved density. 
We demonstrate that such local perturbations (``activations'' or quenches) result in time-dependent density gradients which, in turn, give rise to phoretic mechanisms that can propel inclusions immersed in the fluid \textit{at a distance}, i.e., such influences decay as power laws in space. Our analysis, based on the diffusion equation, holds for any perturbation (``activation'') that couples to a diffusive mode.
We argue below that this is expected to apply to specific fluid media, e.g., colloidal suspensions where particle potentials are perturbed, as well as systems where energy and momentum are not conserved. Examples for the former include cases where particle interactions of the medium can be controlled through external fields~\cite{maretkeim2004} or by rapid changes of the particles' size~\cite{ballauff2006}. Examples for the latter are active fluids~\cite{solon_active_2015, berthier2015epl,catesX,buttinoni2012active,bauerle2018self,loiEffectiveT2008, bizonne, rohwer2018forces} in 2D, where the presence of confining surfaces destroys energy and momentum conservation, and which can be activated locally~\cite{bauerle2018self,santer2016manipulation}. 
Systems of granular matter, such as grains on an air bed~\cite{keys2007measurement,abate2007topological}, vibrated granular matter~\cite{kudrolli2004size}, or self-propelled disks~\cite{dauchot2017disks} are further candidates.

We discuss the basic properties of this ``activated diffusiophoresis'', investigating how it can be used to manipulate inclusions, e.g., by propelling them along well-controlled trajectories. Suitable  activation protocols may also be used to extract work from inclusions which move along closed trajectories in the fluid, so that an  engine cycle can be formulated.
The manuscript is structured as follows. Section~\ref{sec:sys} provides a detailed description and physical examples of the system under consideration, including the basic mathematical model of density dynamics occurring after local activations. The associated forces on an inclusion immersed in the fluid are also discussed. Section~\ref{sec:pulse} deals with the density and forces which emerge after strips within an immersed surface are activated by a travelling pulse. These insights are applied in Sec.~\ref{sec:work}, where we put forward suggestions for an engine: an inclusion moving along a closed trajectory through the evolving post-activation-pulse density field, can perform negative or positive work depending on details of the cycle. Section~\ref{sec:accel} contains the main result of this work: we demonstrate that a protocol of successive activations of adjacent strips within two parallel walls, gives rise to a conveyor, where a traveling density wave transports inclusions along sustained trajectories in the fluid. We analyze the necessary conditions for such stable transport, finding that it is ruled out in steady states, in analogy to Earnshaw's theorem of electroststics \cite{earnshaw1842}, but can be realized when the density changes as a function of time. A dimensionless mobility determines for which scenarios the proposed design is functional. Finally, concluding discussions and an outlook are presented in Sec.~\ref{sec:disc}.

\section{Physical systems and model}
\label{sec:sys}
\begin{figure}[t]
\centering
\includegraphics[width=0.9\columnwidth]{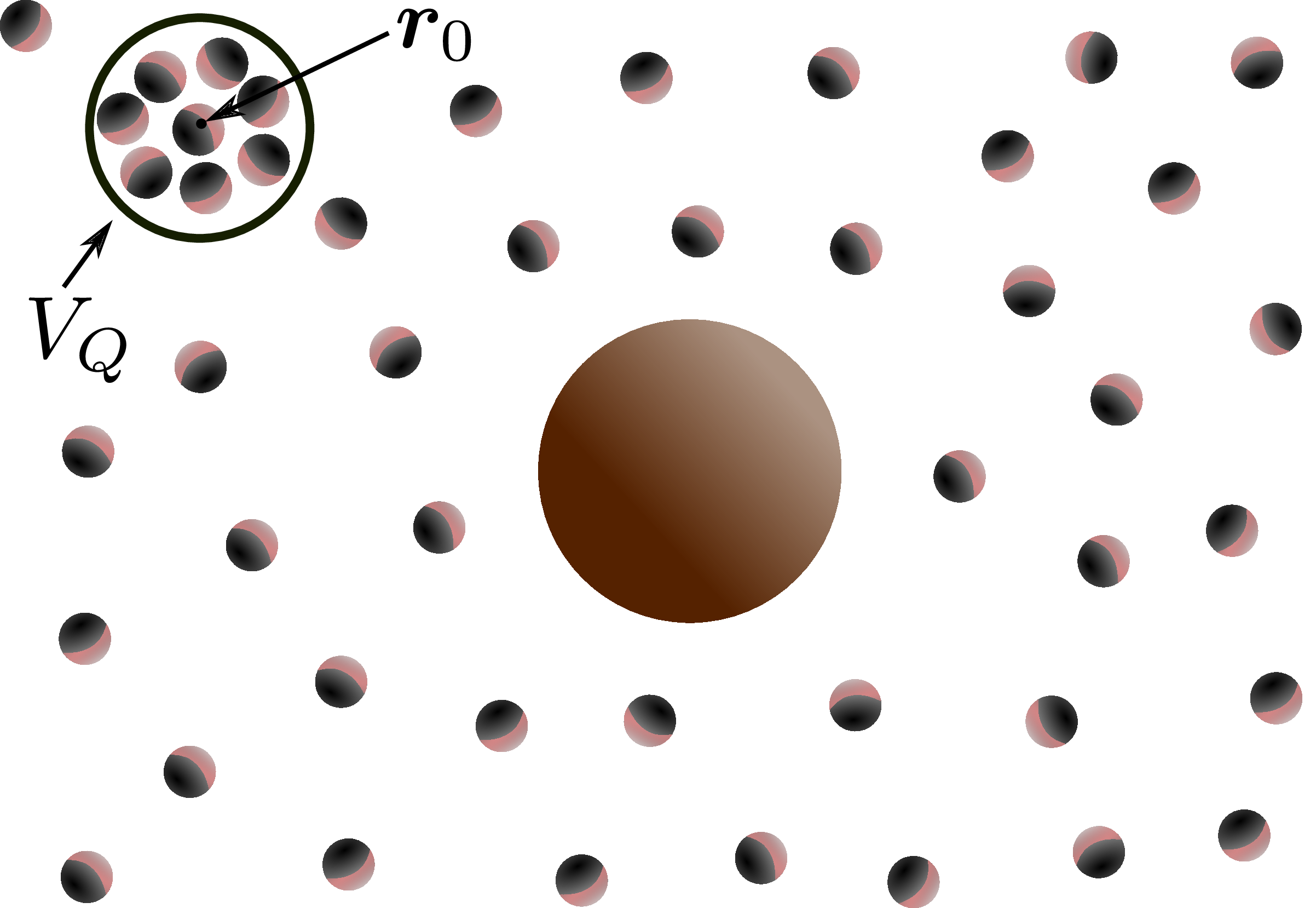}
\caption{Sketch of local activation of a system as described in \Sref{sec:phys}. The activation (e.g., a local change of interaction potential or activity) results in a change of particle density in the volume $V_Q$ at position $\bm r_0$ that propagates diffusively. These activated diffusive waves give rise to phoretic motion of immersed inclusions (large particle). }
\label{fig:fluid}
\end{figure}

\subsection{Physical systems}\label{sec:phys}
The analysis of this manuscript is based on the diffusion equation, and is thus expected to be applicable in a variety of systems. Because the dynamics of non-equilibrium systems or fluids in general depends on the conservation laws involved~\cite{HansenMcDonald}, we provide two specific example of fluid media for which the diffusion equation is expected to capture relevant phenomena.

\begin{enumerate}[label=(\roman*), itemsep=0pt, topsep=0pt]
    \item The first is a suspension of colloidal particles, which, at length scales large compared to particle sizes, is expected to follow a diffusion equation
~\cite{dhont}. This holds also in the presence of hydrodynamic interactions (momentum conservation). An activation can then be applied, e.g., via a change of the interaction potential of the colloidal particles. This has been realized experimentally, e.g., for particles whose interaction strength is controlled via an external magnetic field~\cite{maretkeim2004} or particles whose sizes can  be changed~\cite{ballauff2006}. 
    \item A rather different scenario is presented by systems of active  particles in two spatial dimensions, such as swimmers or vibrated granular systems~\cite{keys2007measurement,abate2007topological,kudrolli2004size,buttinoni2012active,dauchot2017disks}. Even in the presence of a molecular fluid solvent as in Refs.~\cite{buttinoni2012active,bauerle2018self,santer2016manipulation}, the substrate that ``confines'' the system to 2D in all cases intervenes such  that neither momentum nor energy are conserved on large length scales. The remaining conservation law of particle density suggests the use of the diffusion equation. In this case, local activation can be achieved, e.g., by local changes in motility through targeted illumination of Janus-type swimmers in a colloidal suspension~\cite{bauerle2018self}. 
\end{enumerate}
We emphasize that while a 2D system of type (ii) is necessarily constrained by a confining surface, we still refer to it as \textit{bulk}. Later we consider the case where the bulk is additionally confined by no-flux surfaces (or lines for 2D), e.g., in a film geometry [see \Fref{fig:platepulse}]. For clarity, the latter boundaries will be termed \textit{walls}.


\subsection{Activation: basic equations}


Consider a system (e.g., of the type described in \Sref{sec:phys}) at steady state with a homogeneous density $\bar\rho$ initially. Activation of the volume $V_Q$  at position $\bm r_0$ (in a coarse grained description) at time $t_0$, e.g., via the methods described in \Sref{sec:phys} (see also \fref{fig:fluid}) causes a change of particle density in $V_Q$. Thus, after activation, a new coexistence must be established between the bulk system and the volume $V_Q$, via transport of particles  away from or to $V_Q$ (depending on the type of activation). On time scales  large compared to the equilibration time inside $V_Q$, and length scales large relative to the size of this region, this transport process follows the diffusion equation, with an initially increased or decreased local density at $\bm r_0$. We thus have the following initial condition  at time $t=t_0$,
\begin{equation}
\rho(\bm r,t=t_0) = \bar\rho +\bar\rho\alpha\; \delta^{d}(\bm r-\bm r_0). 
\label{eq:rhosplit}
\end{equation}
We introduced the Dirac delta function in $d$ dimensions. 
The initial excess or deficit of particles in $V_Q$ is quantified by the coefficient $\alpha$, which carries units of volume. We expect that $\alpha$ can in many cases be estimated on thermodynamic grounds~\cite{rohwer2018forces}, e.g. from equality of pressure inside and outside $V_Q$.  
As detailed in \Sref{sec:phys}, we consider systems that evolve according to the diffusion equation, so that the density obeys
\al{
\partial_t \rho(\bm r,  t\cmmnt{;\alpha}) = D_0 \nabla^2 \rho(\bm r, t\cmmnt{;\alpha}),
\label{eq:diff}
}
where $D_0$ is the collective small-wavevector diffusion coefficient of the medium~\cite{dhont}. The corresponding Green's function $G$ obeys
\al{
\left(\partial_t  - D_0 \nabla^2\right) G(\bm r, \bm r_0; t,t_0\cmmnt{;\alpha}) =\delta^{d}(\bm r-\bm r_0)\delta(t-t_0).
\label{eq:dif}
}
If several positions or regions are activated at times $t_0({\bm r}_0)$ , we obtain by superposition 
\al{
\rho(\bm r,  t\cmmnt{;\alpha})&=\bar\rho+ \bar\rho\alpha 
\sum_{{\bm r}_0} \theta[t-t_0({\bm r}_0)]
G[\bm r, \bm r_0; t-t_0({\bm r}_0)],
\label{eq:extended}
}
where the summation over various sources becomes an integral for continuous regions. Aiming to study activated diffusiophoresis in different geometries, we provide the corresponding functions $G$ for these cases in the following subsection.

\subsection{Green's functions}
In the bulk (indicated by superscript $b$) and in $d$ spatial dimensions, the solution of Eq.~\eqref{eq:dif} is given by the diffusion kernel~\cite{crank1979,cole2010heat},
\al{
	G\bulk(\bm r, \bm r_0; t,t_0) &=\frac{e^{-(\bm r-\bm r_0)^2/4D_0(t-t_0)}}{[4\pi D_0(t-t_0)]^{d/2}}.
\label{eq:DiffKern}
}
The Green's function $G\bulk$ has the dimension of inverse length to the power $d$.

The Green's function is modified when walls are introduced into the system. We will consider no-flux boundary conditions, i.e., surfaces which cannot be crossed by particles. For the diffusion equation~\eref{eq:dif}, these are described by Neumann boundary conditions~\cite{cole2010heat}, i.e., the density gradient normal to the wall is forced to vanish. In what follows, such walls will be taken to have normal vectors in the $z$ direction, so that $\bm r \equiv (\bm r_\parallel,z)$, with $\bm r_\parallel\in\mathbb R^{d-1}$ spanning the wall. A single wall (indicated by superscript $1w$) at $z=0$ results in the following Green's function, obtained by the method of images~\cite{cole2010heat},
\al{
G\onew(\bm r, \bm r_0; t,t_0) &= G\bulk[\bm r, (\bm r_{0,\parallel},z_0); t,t_0]\nl
&\qquad +G\bulk[\bm r, (\bm r_{0,\parallel},-z_0); t,t_0],
\label{eq:Dens1w}
}
where $\bm r$ and $\bm r_0$ are on the same side of the wall.
For a space confined between two walls (indicated by superscript $2w$) situated at $z= 0,L$, with $0\leq \{z,z_0\}\leq L$, the solution is obtained by using an infinite number of images to ensure that the flux across both walls vanishes~\cite{rohwer2018forces,cole2010heat},
\al{
G\twow(\bm r, \bm r_0; t,t_0) &= \sum_{n=-\infty}^\infty G\onew[\bm r, (\bm r_{0,\parallel},z_0+2n L); t,t_0]. 
\label{eq:Dens2w}
}

\begin{figure}[t]
\centering
\includegraphics[width=0.95\columnwidth]{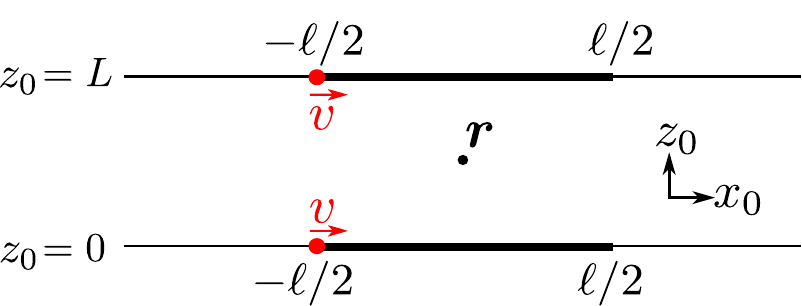}
\caption{
	Top view of a setup of two parallel walls (or lines in 2D) as discussed in \Sref{sec:pulse}.
	Successive points on the two lines are activated, so that a ``pulse'' (red dot) of activation travels along the strip with velocity $v$. The extent of the pulse is restricted to a length $\ell$, so that, in a coordinate frame set by the center of the pulse range, $x_{0M}$,  the pulse travels from $-\ell/2$ to $\ell/2$. This analysis is equivalently applicable to a 3D system that is translationally invariant along the direction pointing into the plane of the sketch.
}
\label{fig:platepulse}
\end{figure}

\begin{widetext}
In what follows, for simplicity, we restrict to two-dimensional problems, where walls reduce to lines. The presented analysis however also holds for 3D systems that are translationally invariant in one direction parallel to walls (here $y$). This means that regions of activation must be translationally invariant as well, so that, instead of points in 2D, lines pointing in the $y$ direction are activated in 3D. In the latter case, \fref{fig:platepulse} represents a top view of the system. We will thus require the 2D propagator for two walls (lines) at $z_0 = 0,L$, for which~\Eref{eq:Dens2w} gives
\al{
G\twow(\bm r, \bm r_0; t,t_0)
&= \frac{e^{-\frac{\left(x-x_0\right){}^2}{4 D_0 \left(t-t_0\right)}} }{L \sqrt{16\pi D_0 \left(t-t_0\right)}}
\left[\vartheta _3\left(-\frac{\pi  \left(z-z_0\right)}{2 L},e^{-\frac{\pi ^2 D_0 \left(t-t_0\right)}{L^2}}\right)+\vartheta _3\left(-\frac{\pi  \left(z+z_0\right)}{2 L},e^{-\frac{\pi ^2 D_0 \left(t-t_0\right)}{L^2}}\right)\right],
\label{eq:Dens2w2d}
}
\end{widetext}
where $\vartheta_3(u,q) = 1+2\sum_{n=1}^\infty q^{n^2}\cos(2nu)$ is the Jacobi elliptic function of the third kind~\cite{abramowitzstegun}.

\subsection{Phoresis: action of density field on inclusions}\label{sec:acdef}
We seek the action of the density field $\rho$ in Eq.~\eqref{eq:extended} on inclusions immersed in the medium. Inclusions are considered to be small compared to their distance from the points of activation, so that they predominantly feel the gradient of $\rho$. 
We discuss two phoretic mechanisms below.

\subsubsection{Case (i): advected inclusion}
When the medium is a suspension [case (i) in Sec.~\ref{sec:phys}], the dynamics of the field $\rho$ directly translates into the motion of an embedded inclusion. Rewriting the diffusion equation \eqref{eq:dif} in terms of the current $\bm J=\bar\rho {\bm v}$,
\al{
\partial_t \rho = -\nabla\cdot \bm J,
\label{eq:dif2}
}
we obtain the local velocity ${\bm v} = -({D_0}{\bar\rho})\nabla \rho$. In the absence of other forces, it is reasonable to assume that the inclusion is advected with the local velocity of the suspension, so that its position $\bm r$ obeys
\begin{align}
\dot{\bm r}= -\frac{D_0}{\bar\rho}\nabla_{\bm r} \rho.
\label{eq:acdif}
\end{align}
This is the trajectory due to activated diffusiophoresis: the inclusion moves passively along the gradient of fluid density $\rho$. 

\subsubsection{Case (ii): inclusion with friction}
Here we discuss case (ii) in Sec.~\ref{sec:phys}, where the inclusion has contact to an aether, which could for example be the confining surface of a 2D active system. We begin by computing the mechanical force acting on the stationary inclusion in terms of its microscopic potential $V(\bm r)$, which captures the inclusion's interactions with particles making up the density field $\rho$. This force is 
\al{
{\vec { F}}(t) &= \int d\bm r  [\nabla_{\bm r} V({\bm r})] \rho(\bm r,t).
\label{eq:Fincl}
}
While formally exact, Eq.~\eqref{eq:Fincl} is difficult to evaluate due to the back-action of the potential $V$ onto the density $\rho$. However, for a spherically symmetric inclusion \footnote{In contrast to symmetric inclusions, asymmetric inclusions immersed in an active medium have been shown to generate currents~\cite{solonkafri2018}.}, the force can be assumed to be proportional to the local density gradient in the above-mentioned coarse graining limit (see, e.g., Refs.~\cite{popescu2016,anderson1989colloid,golestanian2007DesigningSelfPhoretic}), i.e., 
\al{
\bm { F}(\bm r,t) = -\mathcal V \nabla_{\bm r} \rho(\bm r, t).
\label{eq:FinclTD}
}
Importantly, the density $\rho$ in Eq.~\eqref{eq:FinclTD} is the solution of Eq.~\eqref{eq:dif} in the absence of the inclusion. The (generally unknown) coefficient $\mathcal V$ can be given to leading order in $V$ as the ``energetic volume'' $\mathcal V = \int d\bm r\;   V({\bm r})$~\cite{rohwer2018forces}.
Equation~\eqref{eq:FinclTD} is the force acting on an inclusion at rest. If the inclusion moves at a velocity $\dot{\bm r}$ relative to the aether, a friction forces arises. In the absence of inertial effects, the force in Eq.~\eqref{eq:FinclTD} balances that frictional force, yielding the following equation of motion for the inclusion,
\al{
\gamma \dot{\bm r}(t) =  -\mathcal V \nabla_{\bm r} \rho(\bm r, t)\Big|_{\bm r = \bm r(t)},
\label{eq:eom2}
}
where $\gamma$ is a friction coefficient. Note that Eq.~\eqref{eq:eom2} results from the leading behavior in terms of $\dot{\bm r}$ and $\nabla_{\bm r} \rho$. 

\subsubsection{Trajectories of activated diffusiophoresis}
\label{sec:trajgen}
The equations of motion \eqref{eq:acdif} and \eref{eq:eom2} are qualitatively similar, so that the predicted effects of activated diffusiophoresis are qualitatively the same in the different systems mentioned. Combining the cases, we obtain an equation of motion 
\al{\gamma\dot{\bm r} =\bm F(\bm r(t),t),
	\label{eq:trj}
}
in terms of the diffusiophoretic force for the two cases,
\al{
\bm F(\bm r(t),t) &\equiv -\mu 
\nabla_{\bm r} \rho(\bm r, t)\Big|_{\bm r = \bm r(t)},\quad\!
\mu = 
\begin{cases}
\frac{\gamma D_0}{\bar\rho},&\!\!\!\text{(i)}\\
\mathcal V,&\!\!\!\text{(ii)}.\\
\end{cases}
\label{eq:Fdiff}
}
Note that in case i), the prefactor $\gamma$ cancels in Eq.~\eqref{eq:trj}, and the trajectories do not depend on it. Indeed, for  case i) the force $\bm F$ has been introduced as a mathematical construct to facilitate subsequent discussions.

\section{Traveling activation-pulse for a slab geometry}
\label{sec:pulse}

\subsection{Setup and Activation Protocol}


In this section we discuss the geometry of two parallel lines (or walls) subject to a time- and position-dependent activation protocol. In particular, we consider the special case where regions directly on the walls' surfaces are activated in the form of traveling ``pulses'', as in Fig.~\ref{fig:platepulse}. This choice of protocol allows us to discuss basic properties of activated processes, and we also expect it to be experimentally feasible to apply such activations to walls. 
Possible examples include locally altering the interaction potential of the confining wall with the medium (e.g., via optical traps~\cite{neuman2004optical,weinert2009OpticalConveyor} for a colloidal suspension) or by locally modifying the activity of an active medium at the wall (e.g., by varying the illumination intensity at specific positions~\cite{buttinoni2012active,bauerle2018self}, or by providing fuel at the boundary). For granular matter, the two ``walls'' themselves could be moved or deformed slightly, giving rise to an altered density at that position.
We  study the time and space dependence of the phoretic force in Eq.~\eqref{eq:Fdiff}, demonstrating that local activations lead to forces at a distance which exhibit rich dynamics, with temporal decays depending strongly on the position within the system.


Consider the scenario of Fig.~\ref{fig:platepulse}: two walls or lines are placed at $z_0=0,L$ (recalling that, for the case of a 3D system, translational invariance in $y$-direction is implied). The protocol involves activation by a pulse traveling at a velocity $v$ within the parallel-opposed regions $x_0\in[-\ell/2,\ell/2]$ (``strips'') by the two walls.

\begin{figure}[t]
\centering
\includegraphics[width=.99\columnwidth]{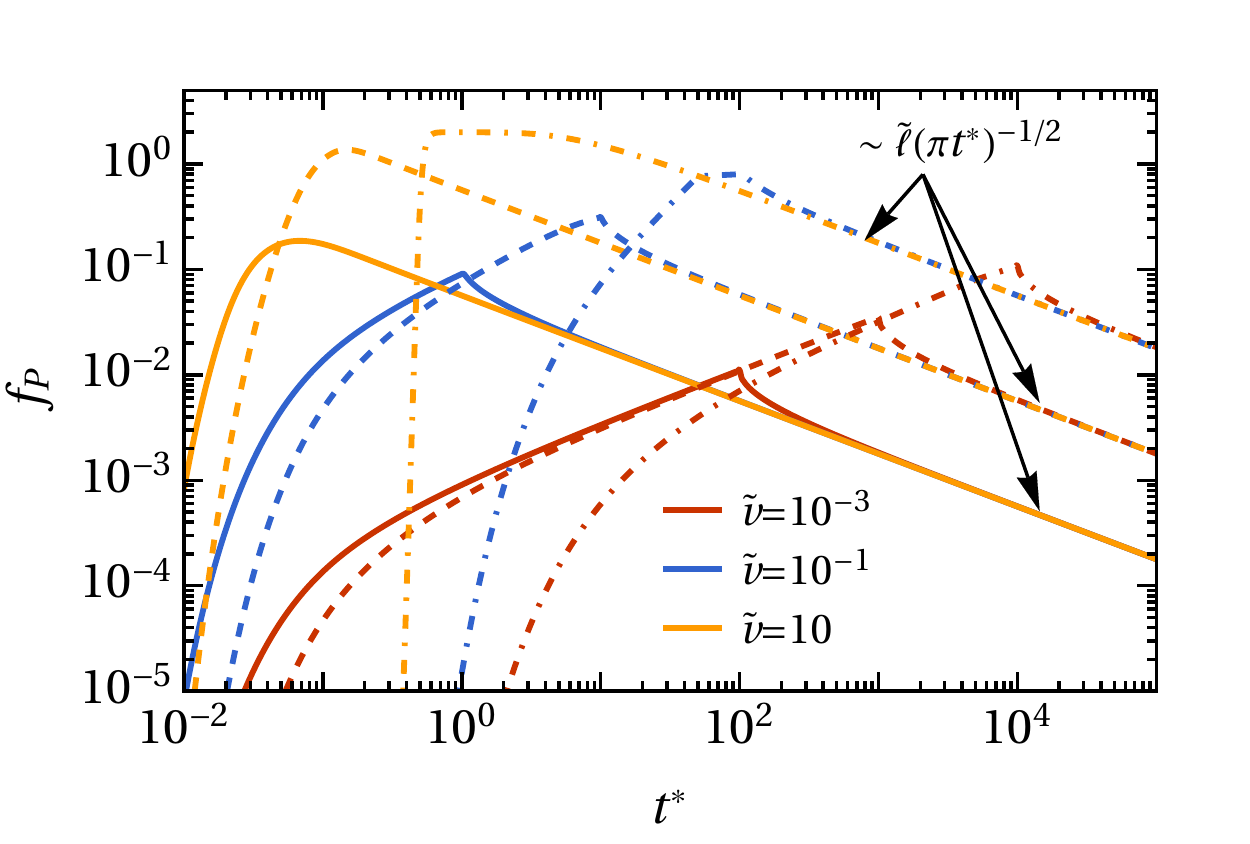}
\caption{Density at the centre of the slab (see \Fref{fig:platepulse}), $\bm r = (x=0,z=L/2)$, as a function of $\ts = D_0 t/L^2$ following a pulse activation traveling the length $\ell$. Plots show the dimensionless density $f_P$ from \Eref{eq:rhopulse2}, where $\rho(x= 0,z=L/2; t;\ell;v;L)=\frac{\tilde \alpha}{L} f_P(\tilde x=0,\tilde z=1/2,\tilde \ell,\tilde v,t^*)$ [see \Eref{eq:rhopulse1}], for $\tilde\ell = \ell/L =$ 0.1 (solid lines), 1 (dashed lines), and 10 (dash-dotted lines), for various values of $\tilde v = vL/D_0$ (see legend).}
\label{fig:densplots}
\end{figure}
\subsection{Density and forces}
Using \Eref{eq:extended}, we find the density field resulting from the activation pulse (indicated by subscript $P$; dependence on $\bm r_0$ and $t_0$ is suppressed),
\al{
\rho_{P}(\bm r, t,\ell,v,L)&= \tilde\alpha \sum_{z_0=0,L} \int_0^{\ell} d x'\; \theta[v(t-t_0)-{x'})] \nl
&\quad \times 
G\twow [\bm r, {\bm r}_0(x'); t,t_0+{x'}/{v}],
\label{eq:rhostrippulse}
}
where $x_{0}(x')= -\ell/2+ x'$ demarcates position within the strips, and
\al{
\tilde\alpha\equiv{\bar\rho\alpha}/{\ell}.
\label{eq:alphatilde}
}
We introduce a parameter $\lambda=1(-1)$ which denotes pulses traveling left to right (right to left), as well as the dimensionless variables \al{\tilde x = \frac x L,\;\; \tilde y = \frac y L,\;\; \tilde \ell = \frac \ell L,\;\; t^* = \frac {D_0 t}{L^2},\;\; \tilde v = \frac{v}{D_0/L},
}
and set the reference time $t_0 = 0$. $\tilde v$ captures the ratio of the pulse speed and the speed of diffusion across a distance $L$, in broad analogy to a P{\'e}clet number. From Eqs.~\eref{eq:Dens2w2d}, \eref{eq:rhostrippulse} 
we find the density at the position $\bm r = (x,z)$ between the walls, 
\al{
\rho_P(\bm r, t,\ell,v,L,\lambda)
&\equiv  \frac{\tilde\alpha}{L} f_P(\tilde x, \tilde z,\tilde \ell,\tilde v,t^*,\lambda), 
\label{eq:rhopulse1}
}
in terms of the dimensionless function
\al{
&f_P(\tilde x, \tilde z,\tilde \ell,\tilde v,t^*,\lambda)\nl
&\quad=\int_0^{\min(\tilde v t^*,\tilde \ell)} d \tilde x'\; \frac{e^{-{\left[\tilde x-\lambda(-\tilde \ell/2+\tilde x')\right]{}^2}/{4 (t^*-\tilde x'/\tilde v)}}}{\sqrt{16 \pi (t^*-\tilde x'/\tilde v)}}  \nl
&\qquad\times \Big(2\vartheta _3 \left[-\frac{\pi}2\tilde z,e^{-\pi ^2 (t^*-\tilde x'/\tilde v)}\right] \nl
&\qquad\quad+ \vartheta _3\left[-\frac{\pi}2(\tilde z-1),e^{-\pi ^2 (t^*-\tilde x'/\tilde v)}\right] \nl 
&\qquad\quad+ \vartheta _3\left[-\frac{\pi}2(\tilde z+1),e^{-\pi ^2 (t^*-\tilde x'/\tilde v)}\right] \Big),
\label{eq:rhopulse2}
}
which can be evaluated numerically. \Fref{fig:densplots} shows the solution for the dimensionless density at the geometric midpoint $(x=0,z=L/2)$ as a function of time after starting the pulse. Initially, $f_P$ grows (from an essential singularity at $t^* = 0$) as long as the pulse is still moving. If $\tilde \ell=\ell/L$ is sufficiently small, there is an additional regime where the density grows algebraically in time, followed by a sharp maximum at $t^* = \tilde \ell/\tilde v$. While some curves seem to display a mathematical kink, the curvature stays finite at this point. Expanding the integrand in \Eref{eq:rhopulse2} for large times shows that the density decays algebraically,  
\al{
	f_P(0, 1/2,\tilde \ell,\tilde v,t^*\gg1) \to \frac{\lt}{\sqrt{\pi\ts}}.
}
The larger the value of $\tilde v$ (more rapid pulses), the more rapidly the density grows, as expected. For a given $\tilde\ell$, $\tilde v$ also affects the maximum of the density, due to the interplay of excess fluid provided by activation, and fluid transport through diffusion.

\begin{figure}[t]
\centering
\includegraphics[width=0.99\columnwidth]{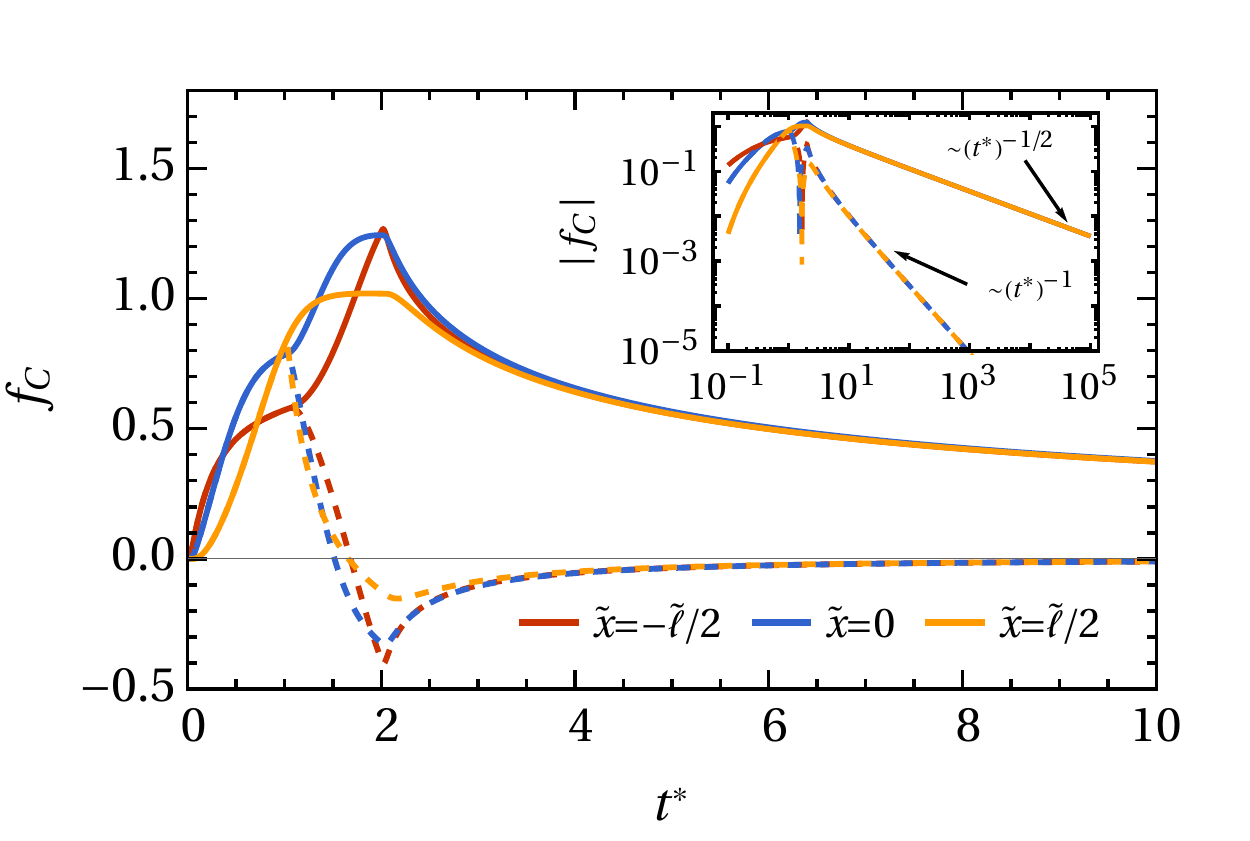}
\caption{Density $\rho\cyc$ along the central line of the slab, $z = L/2$, for various $\tilde x=x/L$ (see legend) as a function of $\ts = D_0 t/L^2$, for a cycled pulse. Shown is $f\cyc$, where $\rho\cyc(x,z=L/2; t;\ell;v;L)=\frac{\alphat}{L} f\cyc(\tilde x,\tilde z=1/2,\tilde \ell,\tilde v,t^*)$ [see \Eref{eq:rhocyc}], for $\tilde\ell =\ell/L =1$ and $\tilde v={v}L/{D_0}=1$. Solid lines represent a $++$ cycle; dashed lines represent a $+-$ cycle.}
\label{fig:densplotscycle}
\end{figure}

The phoretic force of Eq.~\eqref{eq:Fdiff}, acting on an inclusion at a time-independent position, can now be studied  via Eqs.~\eref{eq:rhopulse1} and \eref{eq:rhopulse2}. For an inclusion placed on the central line ($z=L/2$) of the slab, the force points along the $x$-direction due to symmetry, and \Eref{eq:Fdiff} yields
\al{
\bm F_P(x,z= L/2,t) &= -\bm e_x 
\mu \partial_x \rho_P(x,z= L/2, t)\nl
&\equiv \bm e_x \frac{\mu 
\tilde\alpha}{L^{2}} \; g_P(\tilde x, \tilde \ell, \tilde v, t^*,\lambda),
\label{eq:Fgen1}
}
where
\al{
&g_P
= \int_0^{\min(\tilde v t^*,\tilde \ell)} d \tilde x'\; e^{-{\left[\tilde x-\lambda(-\tilde \ell/2+\tilde x')\right]{}^2}/{4 (t^*-\tilde x'/\tilde v)}}  \nl
&\times \left[\frac{\lambda(\tilde \ell/2+\tilde x - \tilde x')}{8\sqrt{\pi }(t^* - \tilde x' / \tilde v)^{3/2}} \right]
\Bigg[2\vartheta _3 \big(-\frac{\pi}4,e^{-\pi ^2 (t^*-\tilde x'/\tilde v)}\big)+\nl
& \vartheta _3\big(\frac{\pi}4,e^{-\pi ^2 (t^*-\tilde x'/\tilde v)}\big) + \vartheta _3 \big(-\!\frac{3\pi}4,e^{-\pi ^2 (t^*-\tilde x'/\tilde v)}\big)\Bigg].
\label{eq:Fgen2}
}
This force will be studied in more detail for the case of cycled pulses (consecutive pulses traveling in opposite directions) in the following subsection.
\subsection{Cycled pulses}

We designate a cycled pulse as one traveling (at each wall) along the $x$ axis from $-\ell/2\to\ell/2\to-\ell/2$. 
By changing activation protocol we introduce two different types of cycling:
In a $++$ cycle, $\alpha$ has equal signs for both directions (i.e., resulting in a net activation of $2\alpha$), while for a $+-$ cycle the reverse pulse has the opposite sign (i.e., after one cycle, the activation status is reversed). The $+-$ cycle thus ultimately brings the system back to the original steady state. The density for such a cycled pulse, indicated by subscript $C$ [we suppress the dependence on $x$, $y$, $L$, $\ell$, and $v$ for brevity; see Eqs.~\eref{eq:rhopulse1} and \eref{eq:rhopulse2}], is  
\al{
\rho\cyc(
t,
\sigma)
&=  \frac{\alphat}{L} \Big[ f_P(t^*,
\lambda=1) 
\nl
& \qquad+ \sigma\theta(t^*-\lt/\vt) f_P(
t^*-\lt/\vt,\lambda=-1)\Big]\nl
&\equiv \frac{\alphat}{L} f\cyc(\tilde x, \tilde z=1/2,\tilde \ell,\tilde v,t^*),
\label{eq:rhocyc}
}
with $\sigma = 1$ or $-1$ for $++$ or $+-$ cycles, respectively.

\begin{figure}[t]
\centering
\includegraphics[width=0.99\columnwidth]{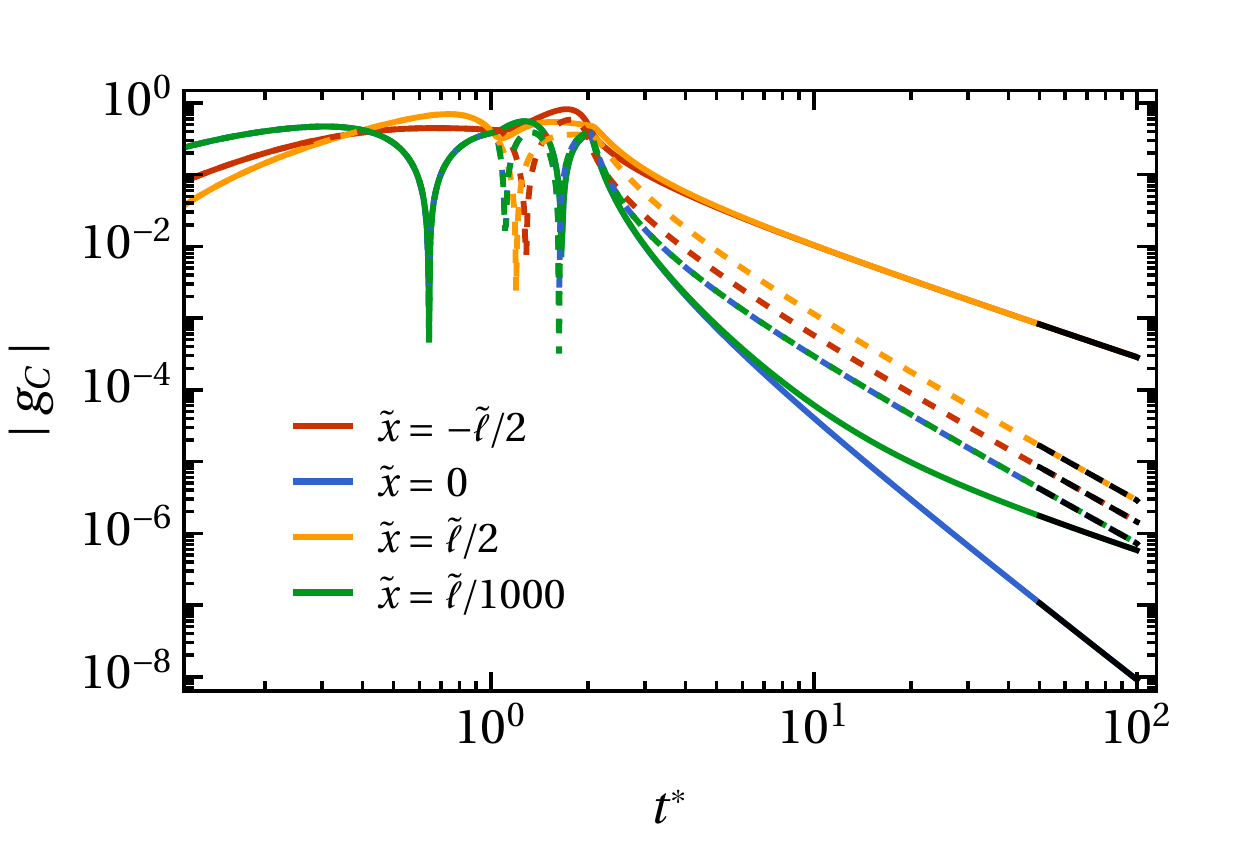}
\caption{The force $\bm F\cyc$ on a small inclusion at various positions $\tilde x=x/L$ (see legend) along the central line of the slab, $z = L/2$, as a function of $\ts = D_0 t/L^2$, for a cycled pulse. Shown is $g\cyc$, where $\bm F\cyc(x,z=L/2; t;\ell;v;L)=\bm e_x (\mathcal V \alphat/L^2) \; g\cyc(\tilde x, \tilde \ell, \tilde v, t^*)$ [see \Eref{eq:Fcyc}], for $\tilde\ell = \ell/L = 1$ and $\tilde v={v}L/{D_0}=1$. Solid lines represent a $++$ cycle; dashed lines represent a $+-$ cycle. The black lines at late times are the exact asymptotes given in \Eref{eq:FcycleStatAsymp}.
}
\label{fig:forcecycle}
\end{figure}

\Fref{fig:densplotscycle} shows the density at various points along the central line $z=L/2$ of the slab. At early times, the dynamics depends on the $\tilde x$ coordinate, since the position relative to the pulse matters. Late-time dynamics is more universal: For $++$ cycles the density decays as $\sim (t^*)^{-1/2}$, as was the case for a single pulse. In contrast, for a $+-$ cycle, leading contributions cancel as expected, resulting in a faster decay $\sim (t^*)^{-1}$. These subtleties also enter the force corresponding to the density in \Eref{eq:rhocyc},
\al{
\bm F\cyc(x,z= L/2,t) &= -\bm e_x 
\mu \partial_x \rho\cyc(x,z= L/2, t)\nl
&\equiv \bm e_x \frac{
\mu \alphat}{L^2} \; g\cyc(\tilde x, \tilde \ell, \tilde v, t^*), 
\label{eq:Fcyc}
}
where $g\cyc = (1/L)\partial_{\xt} f\cyc$ follows from Eqs.~\eref{eq:Fgen1}, \eref{eq:Fgen2} and \eref{eq:rhocyc}. The late-time limit, $\ts\to\infty$, can be given in an asymptotic expansion,
\begin{widetext}
\al{
\bm F\cyc(x,z=\frac L 2,\sigma,t\to\infty) &= \bm e_x \frac{
\mu \alphat}{L^2} \Bigg[ 
\frac{(\sigma +1) \xt}{2 \sqrt{\pi } \left(t^*\right)^{3/2}} 
+ \frac{-\lt^2 (\sigma  (\vt \xt-2)+\vt \xt+2)+12 \lt (3 \sigma  \xt+\xt)-4 (\sigma +1) \vt \xt^3}{32 \sqrt{\pi } \left(t^*\right)^{5/2} \vt}\nl
& \qquad + \frac{1}{3072 \sqrt{\pi } \left(t^*\right)^{7/2} \vt^2}
\Big\{ 3 \lt^4 \sigma  \vt^2 \xt+3 \lt^4 \vt^2 \xt-12 \lt^4 \sigma  \vt+12 \lt^4 \vt+720 \lt^3 \sigma -360 \lt^3 \sigma  \vt \xt-120 \lt^3 \vt \xt \nl
& \qquad -240 \lt^3+40 \lt^2 \sigma  \vt^2 \xt^3+40 \lt^2 \vt^2 \xt^3-240 \lt^2 \sigma  \vt \xt^2+240 \lt^2 \vt \xt^2+6720 \lt^2 \sigma  \xt+960 \lt^2 \xt \nl
& \qquad -1440 \lt \sigma  \vt \xt^3-480 \lt \vt \xt^3+48 \sigma  \vt^2 \xt^5+48 \vt^2 \xt^5\Big\} + \mathcal{O}\tsb^{-9/2}
\Bigg].
\label{eq:FcycleStatAsymp}
}
\end{widetext}
Recall that $\sigma = 1(-1)$ for $++(+-)$ pulses. \Eref{eq:FcycleStatAsymp} thus illustrates that the force depends strongly on the position $\xt$ of the inclusion as well as details of the activation protocol ($\sigma$ and $\vt$). The late-time behavior displays a number of subtleties, as is seen in \Fref{fig:forcecycle}, which shows the force at various points on the central line $z=L/2$ for $\lt=\vt=1$. For a $+-$ cycle, the force decays as $\sim (t^*)^{-5/2}$ independent of $\xt$. In contrast, for a $++$ cycle, the late-time decay shows a singularity at $\xt=0$: For any finite $\xt$ (\Fref{fig:forcecycle} shows $\tilde x = \pm \tilde\ell/2$), the force decays as $\sim (t^*)^{-3/2}$, whereas,  for  $\tilde x= 0$ ,the force decays more rapidly, as $\sim (t^*)^{-7/2}$. This observation is due to a cancellation of leading orders, 
and shows that the limits of small $\xt$ and large $t^*$ do not commute. The green curve in the figure, showing $\xt=\lt/1000$ demonstrates the crossover from $\sim (t^*)^{-7/2}$ to $\sim (t^*)^{-3/2}$ at late times.

As stated, a $+-$ pulse reverts the activation, ultimately returning the fluid to its original steady state. Due to our analysis being linear in activation $\alpha$, one may convince oneself that the time integral of the corresponding force vanishes. Indeed, we showed numerically (within numerical error) 
\footnote{We used interpolation of numerical data for early times together with the exact late time asymptotes of \Eref{eq:FcycleStatAsymp}.} 
that
\al{
\int_0^\infty  d t \; \bm F\cyc(x,z=L/ 2,\sigma=-1,t) = 0.
\label{eq:Fintegral}
}
More generally, this result can be proved for any coordinates $x$ and $z$. Writing Eq.~\eqref{eq:diff} in Laplace space for time, with an initial (activated) density $\delta\rho_0$, yields
\begin{align}
	-\delta\rho_0(\bm r)+s \delta\rho(\bm r,  s\cmmnt{;\alpha}) = D_0 \nabla^2 \delta\rho(\bm r, s\cmmnt{;\alpha}).
\end{align}
Because $\delta\rho(\bm r,  t\cmmnt{;\alpha})$ decays to zero at large times (as the system is unbounded), we have $\lim_{s\to0}s \delta\rho(\bm r,  s\cmmnt{;\alpha})=0$, so that $\delta\rho(\bm r, s=0\cmmnt{;\alpha})$ obeys the Poisson equation of electrostatics with charge density $\frac{\epsilon_0}{D_0}\delta\rho_0(\bm r)$ (introducing $\epsilon_0$, the permeability of vacuum). After a $+-$ cycle, the density $\delta\rho(\bm r, s=0\cmmnt{;\alpha})$ thus obeys the Poisson equation of charge-free space. Further, $\delta\rho(\bm r, s=0\cmmnt{;\alpha})$ vanishes at the walls, and $\lim_{x\to\pm\infty}\delta\rho(\bm r, s=0\cmmnt{;\alpha})=0$, so that the solution is $\delta\rho(\bm r, s=0\cmmnt{;\alpha})=0$. For a $+-$ cycle, the time integral of the phoretic force  thus  vanishes at every point, i.e., this cycle results in a zero net force on stationary inclusions. This resembles the famous problem of swimming at low Reynolds numbers~\cite{purcell1977LowRe,najafi2005PropLowRe,lauga2009HydrodynSwim}, where time-asymmetric mechanical motion is required in order to generate a net displacement of swimmers. We shall show in \Sref{sec:accel} below, that inclusions which follow the trajectories of Eq.~\eqref{eq:trj} can, however, be transported by cyclic protocols.

\section{Work extracted by a moving inclusion}
\label{sec:work}

\begin{figure}[t]
\centering
\includegraphics[width=0.99\columnwidth]{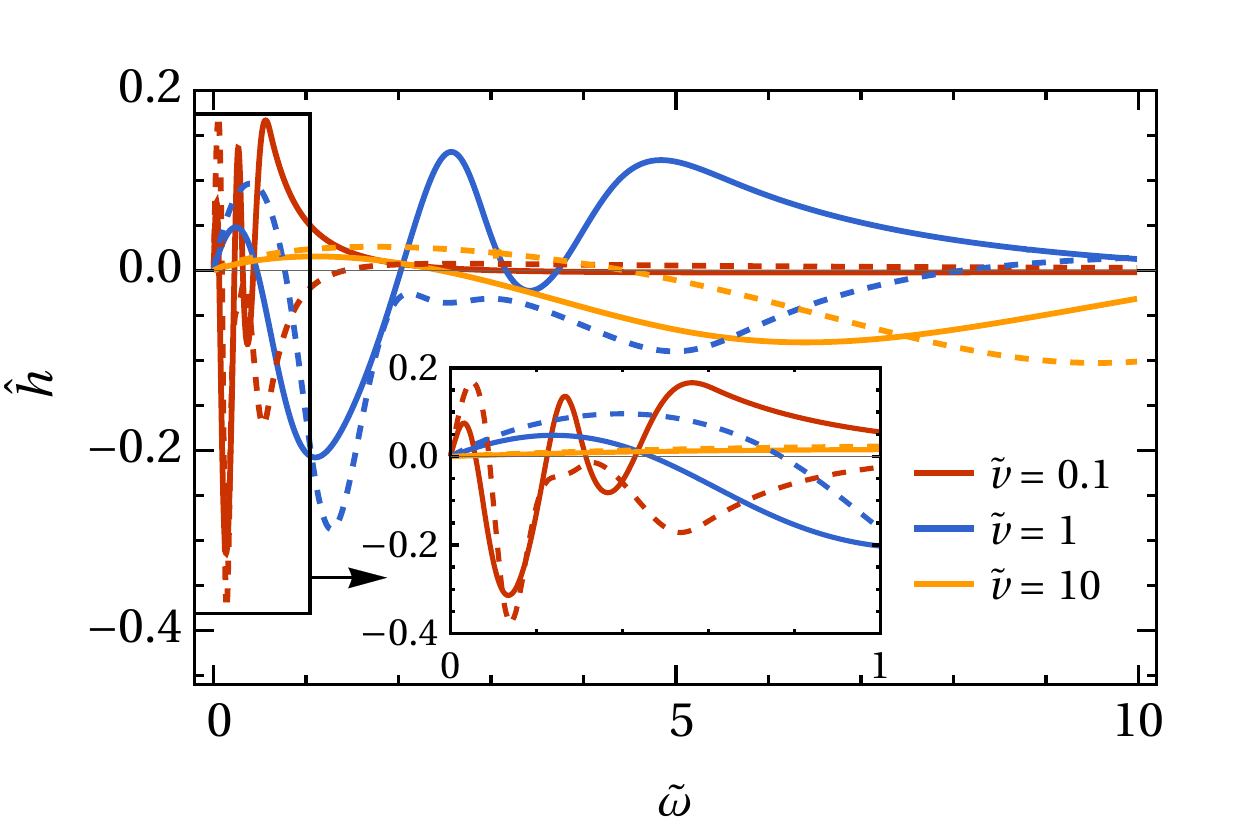}
\caption{
The ratio $W/W_D = \mut \hat h$ of the work against density gradients $W$ and the dissipated work $W_D$ [see \Eref{eq:workRatio}], exerted by an inclusion during one period of a closed trajectory $\tilde X(t) = -\tilde A \cos(\tilde\omega t^*)$ in the presence of a cycled pulse [see \Eref{eq:Fcyc}]. Here $\mut$ is the effective mobility [\Eref{eq:muDef}], $\lt =\ell/L= 1$, $\zt=z/L=1/2$ and $\tilde A =\lt/2$ (solid lines) or $\tilde A =-\lt/2$ (dashed lines), $\ts = D_0 t/L^2$, $\vt={v}L/{D_0}$ and $\tilde\omega = \omega L^2/D_0$.
}
\label{fig:WorkClosedTraj}
\end{figure}

Microscopic heat engines that rectify thermal fluctuations by periodic motion of a colloid have been realized experimentally, see Refs.~\cite{blickle2012MicroHeatEngine,martinez2017ColloidalEngine}. Here we propose a similar notion with an inclusion  exposed to dynamic density gradients. Consider an inclusion which is mounted on an external arm that moves it on a trajectory $\bm X(t)$. As analyzed in Sec.~\ref{sec:acdef}, the arm is force-free if $\dot{\bm X}(t)=\dot{\bm r}(t)$, where ${\bm r}(t)$ is the particle trajectory given in Eq.~\eqref{eq:trj}. The force acting on the arm is thus proportional to the deviation of $\dot{\bm X}(t)$ from $\dot{\bm r}(t)$. Therefore the work $W$ done by the arm is
\al{
W =  \int_0^\tau dt\; \gamma\left(\dot{\bm X}(t)  + \frac{\mu}{\gamma}\nabla \rho \right) \cdot \dot{\bm X}(t),
}
with $\mu$ given in \Eref{eq:Fdiff} for the advection and aether friction models, respectively.

We consider a closed trajectory along the central line of the slab: $Z=L/2$ and $X(t) = -A \cos(\omega t)$, i.e., $\tilde X(t) = -\tilde A \cos(\tilde\omega t^*)$, where $\tilde\omega = \omega L^2/D_0$ is a dimensionless frequency and $(\tilde A,\tilde X)=(A/L,X/L)$. For a cycled activity pulse with velocity $v$ and signature $+-$, as discussed in detail in \Sref{sec:pulse} [see Eqs.~\eref{eq:rhocyc} and \eref{eq:Fcyc}], the work is
%
\al{
	W  = W_D \left(1-\mut \hat h [\tilde A, \tilde \ell, \tilde v, \tilde\omega ]\right),
\label{eq:workRatio}
}
where parameter dependencies have been reduced into a dimensionless mobility coefficient
\al{
\tilde\mu &= \frac{\mu \alphat}{\gamma D_0 L}.
\label{eq:muDef}
}
$W_D$ is the work done in the absence of activation against the dissipative force,  
\al{
W_D = \int_0^\tau dt\; \gamma\dot X^2 = \gamma A^2 \pi\omega = \gamma \pi \tilde A^2\omegat\geq 0.
\label{eq:Wdiss}
}
The reduced mobility in Eq.~\eqref{eq:muDef} depends on various parameters such as the size of the inclusion, the magnitude of the density wave induced by the activation, the friction coefficient of the inclusion, the diffusion coefficient of the medium, and the wall separation. Figure~\ref{fig:WorkClosedTraj} shows the function $\hat h$ for various parameter combinations. For rapid pulses (e.g., $\vt=10$), the 
asymmetry of the pulse relaxes rapidly and the inclusion sees an essentially symmetric density (particularly for small frequencies), so that reversing the trajectory (interchanging $A\to-A$ or, equivalently, changing the direction of the pulse) does not change the sign of the performed work. For slower pulses (e.g., $\vt=0.1$ or 1), $\hat h$ can change its sign depending on $\omegat$ and on whether $A\lessgtr0$, since the  frequency and initial position determine which density gradients the particle is exposed to during its trajectory. At smaller frequencies $\omegat$, details of $W$ depend strongly on $\omegat$ and $A$ for slower pulses. 

Whenever $W<0$ in Eq.~\eqref{eq:workRatio}, which can be achieved for any value of $\hat h$ by adjusting $\tilde \mu$ (including its sign via the sign of $\tilde \alpha$ or $\mu$), the external arm extracts work from the system. For the given $+-$ cycle, this process can be repeated (after both the position $\bm X$ and the activation pulse have reached their initial positions), so that the system acts as an engine. The rich phenomenology described  in Fig.~\ref{fig:WorkClosedTraj} could thus be exploited to obtain various engine modes.

\section{A Conveyor}
\label{sec:accel}


\begin{figure}[t]
\centering
\includegraphics[width=0.99\columnwidth]{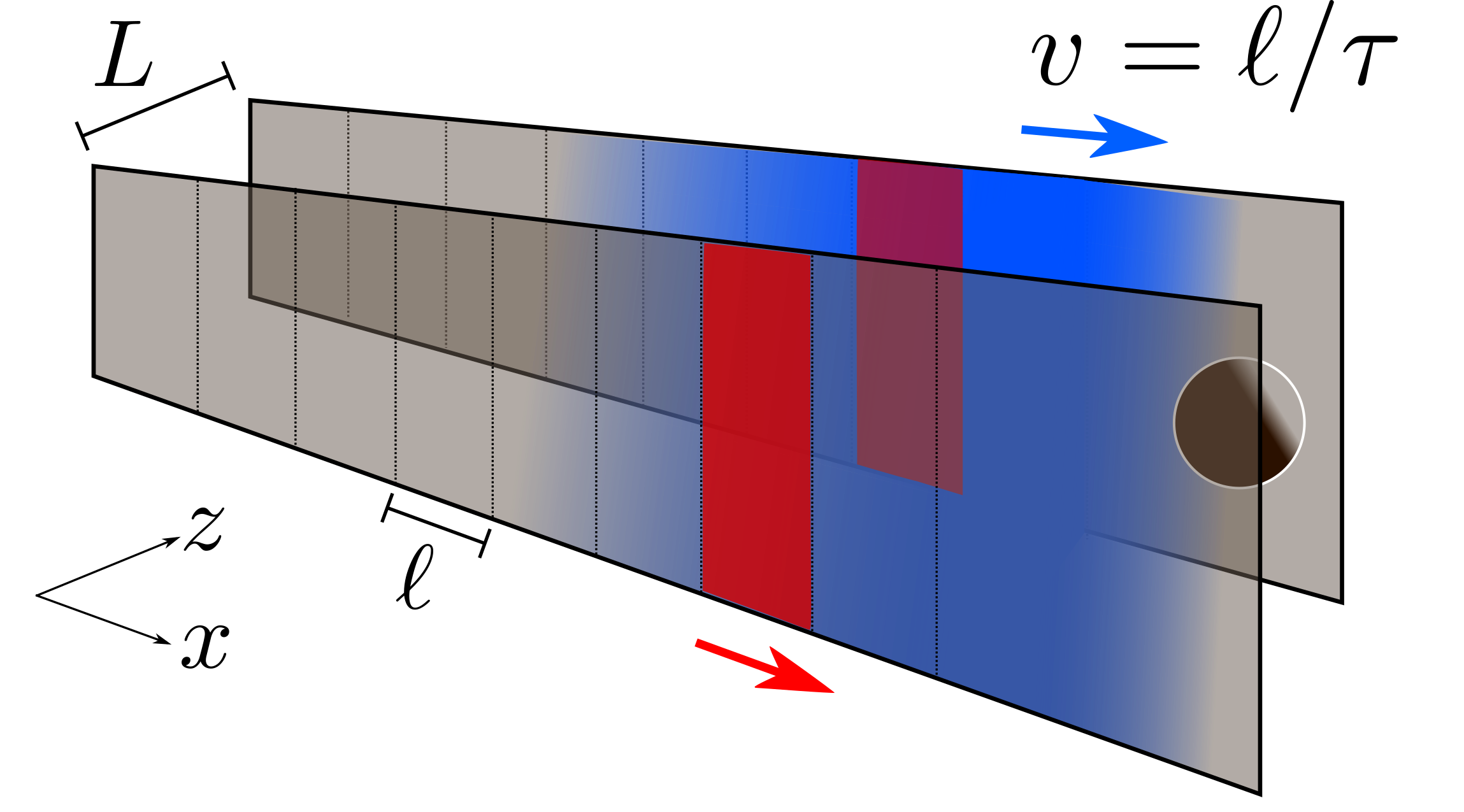}
\caption{
	Schematic of the conveyer analyzed in Sec.~\ref{sec:accel}. Pairs of regions (``strips''), of width $\ell$ each, are activated sequentially, left to right, with a delay time $\tau$ between each activation. This yields a steady state with a density wave (of complex shape) traveling at an average velocity $\bm v=(\ell/\tau) \bm e_x$ from left to right. This wave can convey inclusions on trajectories which are stable against small perturbations, if the inclusions' mobility is large enough. 
}
\label{fig:accel}
\end{figure}

In a state with time-independent density, Eq.~\eqref{eq:diff} yields $\nabla^2\rho=0$. The diffusiophoretic force in Eq.~\eqref{eq:Fdiff}, $\vec{F}=-\mu\nabla\rho$, is thus divergence-free at any point, $\nabla\cdot\vec{F}=0$, in steady state. Therefore the force field $\vec{F}$ has no sinks, so that no stable positions (of zero force) can exist for inclusions, even in presence of time-independent currents. Pointing once more to the similarity to electrostatics, we note that the phoretic force maps onto the electrostatic force for a potential $\rho$, illuminating the similarity to Earnshaw's theorem~\cite{earnshaw1842}. 

In contrast, in regions where the density increases in time, $d\rho/dt=D_0\nabla^2\rho>0$, stable points may exist. 

In this section we exploit this insight, and present a method of harnessing successive activations in order to generate persistent trajectories of inclusions in a setup confined between two  parallel walls (or lines in 2D). The employed protocol is a variant of the protocol used on \Sref{sec:pulse}, where the range of pulse is extended to infinity, and the pulse velocity is discretized for computational ease. Practically, this protocol is achieved by adding an infinite amount of ranges (``strips''), and activating each range instantaneously. A schematic of the conveyor design is shown in Fig.~\ref{fig:accel}.

\subsection{Activating individual strips instantaneously: density and forces}
\label{sec:InstDensAndF}
A first step towards the protocol of conveyor is instantaneous activation of a strip of lentgh $\ell$ on the two walls, see also Fig.~\ref{fig:accel}. The time-dependent density for this case 
is obtained by setting $v\to\infty$ in Eqs.~\eref{eq:rhopulse1} and \eref{eq:rhopulse2}, 
which renders the $x'$ integral analytically tractable. For the two parallel opposed strips [see \Fref{fig:platepulse} (b)], whose midpoint is situated at $x_{0M} = 0$, the density following an instantaneous activation (indicated by subscript $I$) is 
\al{
\rho_I(\bm r; t;\ell;L|x_{0M}=0)&=\rho_P(\bm r; t;\ell;v\to\infty;L)\nl
&\equiv  \frac{\tilde\alpha}{L} f_I(\tilde x, \tilde z,\tilde \ell,t^*),
\label{eq:rhoinst1}
}
with
\al{
&f_I(\tilde x,\tilde z, \tilde \ell, t^*) = \nl
&\frac 1 4
\Bigg[\text{erf}\big(\frac{2 \tilde{x}+\tilde{\ell }}{4 \sqrt{t^*}}\big)-\text{erf}\big(\frac{2 \tilde{x}-\tilde{\ell }}{4 \sqrt{t^*}}\big)\Bigg] 
\Bigg[
2\vartheta _3\big(-\frac{\pi\tilde z}2 ,e^{-\pi ^2 t^*}\big) +\nl
& \vartheta _3\big(\frac{\pi(1-\tilde z)}2,e^{-\pi ^2 t^*}\big)
+ \vartheta _3\big(-\frac{\pi(\tilde z+1)}2,e^{-\pi ^2 t^*}\big)
\Bigg].
\label{eq:rhoinst2}
}
As before, the phoretic force in Eq.~\eqref{eq:Fdiff} follows from Eqs.~\eref{eq:rhoinst1} and \eref{eq:rhoinst2}. 
For brevity we only provide the force for a particle situated along the central line ($\tilde z=1/2$) of the slab,
\al{
\bm F_{I} (x,z=L/2,\sigma,t) &= \bm e_x\frac{
 \mu\alphat}{L^2} g_I(\tilde x, \tilde \ell, t^*),
\label{eq:Finst1}
}
where
\al{
g_I(\tilde x, \tilde \ell, t^*) &= 
\frac 1 {4 \sqrt{\pi t^*}}
\Big[e^{-{\left(\tilde{\ell }-2 \tilde{x}\right)^2}/{16 t^*}}-e^{-{\left(2 \tilde{x}+\tilde{\ell }\right)^2}/{16 t^*}}\Big] \nl
&\quad\times\Big[\vartheta _3({\pi }/{4},e^{-\pi ^2 t^*})+2 \vartheta _3(-{\pi }/{4},e^{-\pi ^2 t^*})\nl
&\qquad+\vartheta _3(-{3\pi }/{4} ,e^{-\pi ^2 t^*})\Big].
\label{eq:Finst2}
}

In Appendix \ref{app:AppStripQuench} we discuss in detail the density and its gradient, given in Eqs.~\eqref{eq:rhoinst2} and \eqref{eq:Finst2}, respectively. By symmetry, the force $F_x$ parallel to the walls is an asymmetric function of $x$, and the perpendicular force $F_z$ is zero along the central line $\tilde z=1/2$. In the vicinity of the central line, the force $F_z$ points towards the central line, rendering it stable in the sense that inclusions are pushed towards the center. The force $F_z$ however decays rapidly as a function of time, on a time scale of roughly $\ts\approx 10^{-1}$, because this is the time scale of homogenization of density along $z$ (\Fref{fig:rho}). For larger times, the phoretic force therefore points in the $x$ direction, independent of $z$ (see \Fref{fig:Fx}).

\subsection{Conveyor setup and trajectories}

As sketched in \Fref{fig:accel}, we aim to create a steady motion of inclusions through successive activation of strips (of length $\lt$), so that the $n$-th pair is activated at $t^* = n\taus=nD_0 \tau/L^2$. This leads to a superposition (indicated by subscript $S$) of densities of the form in Eq.~\eref{eq:rhoinst1},
\al{
\rho_S(\bm r;t) = \sum_{n=-\infty}^{\text{Floor}(t/\tau)} \rho_I(\bm r; t+n \tau|\xt_{0M}=n\lt),
\label{eq:rhoaccel}
}
where $n=-\infty$ implies that all transient effects after initiation of a first region have ceased, and a steady wave front approaches from the left at a speed $\vt_{\textrm{s.s.}}=\lt/\taus$. As this wave approaches, the density increases, and, according to the arguments above, a stable position can exist when the friction force is added to the phoretic force. In other words, for a wave moving at speed $\vec{v}$,  $D_0\nabla^2\rho=-\bm v\cdot \nabla\rho$. An inclusion being carried along by such a wave has the velocity $\bm v=-{\mu/\gamma}\nabla\rho$, and thus experiences an effective potential that (in the moving frame) satisfies $\nabla^2\rho={\mu/\gamma}(\nabla\rho)^2>0$. It is therefore in principle possible to convey an inclusion stably via the forces emerging from a diffusive wave.  


\begin{figure}[t]
\centering
\includegraphics[width=0.99\columnwidth]{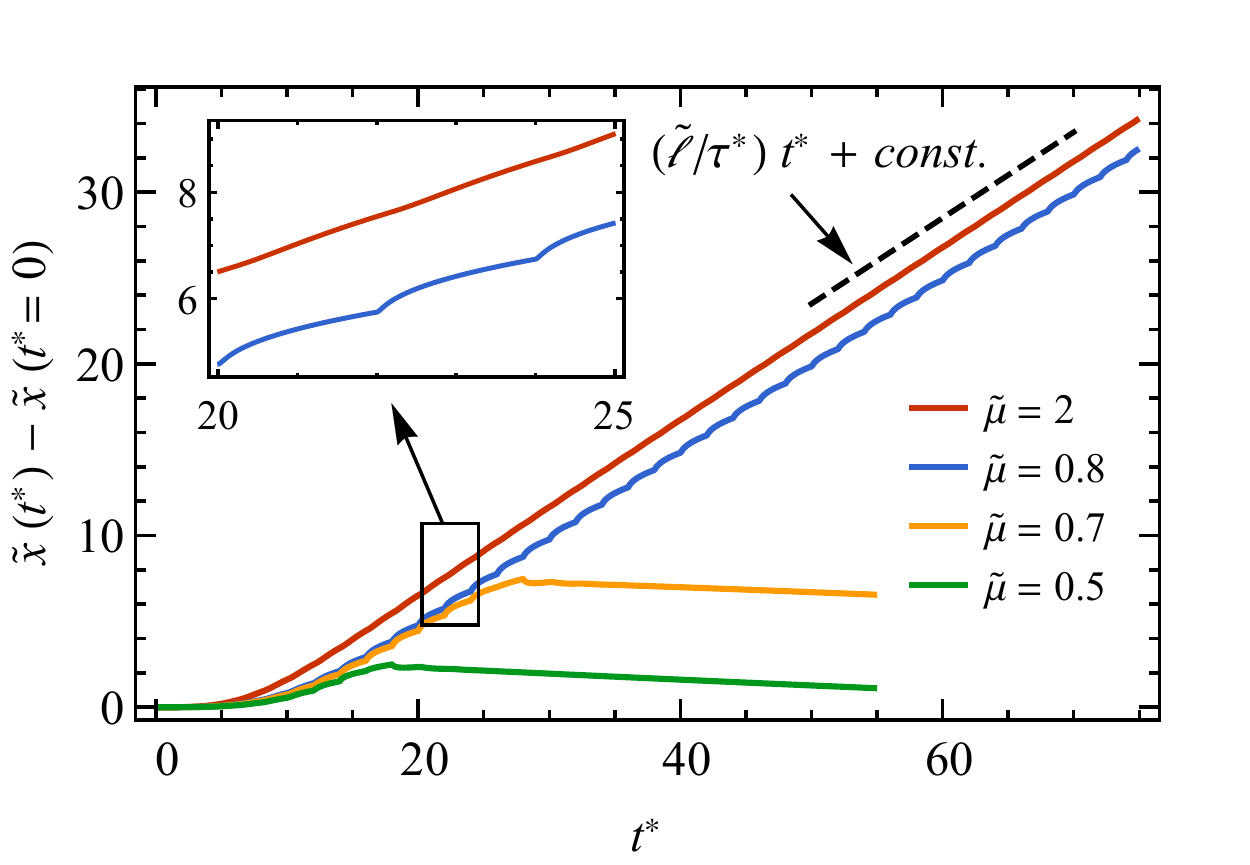}
\caption{Trajectories of an inclusion that is impinged upon by a steady-state wave-front from the left, for various effective mobilities $\mut={\mu \alphat}/{\gamma D_0 L}$. Here $\xt=x/L$, $\zt=z/L=1/2$, $\ts = D_0 t/L^2$, $\lt=\ell/L = 1$, and $\taus =D_0 \tau/L^2= 2$, i.e., the steady state velocity is $\vt_{\textrm{s.s.}}=\lt/\taus=1/2$.
}
\label{fig:trajectories}
\end{figure}


At time $t=-\infty$ an inclusion is placed at the position $(\xt=0,\zt=1/2)$; the inclusion is then impinged upon by the density wave in Eq.~\eqref{eq:rhoaccel}. Its trajectory, which follows Eq.~\eqref{eq:trj}, is found numerically by Euler integration, and is shown in \Fref{fig:trajectories} for various values of $\tilde \mu$ [defined in \Eref{eq:muDef}]. 
For any $\mut$, the inclusion will experience some initial displacement (the direction of which depends on the sign of $\mu$). However, only sufficiently large $\mut$ lead to sustained trajectories (red and blue curves in \Fref{fig:trajectories}). Due to the discrete nature of strip activation, the trajectories display a wavy form, with an average or coarse-grained speed of $\vt$. If $\mut$ is very large, the particle remains far ahead of the density wave, and the wavy character disappears (red curve).
If $\mut$ is too small (e.g., the yellow or green curves in \Fref{fig:trajectories}), the inclusion is at some stage ``overtaken'' by the density front, and eventually comes to rest ---  a ``failed'' trajectory.
\begin{figure}[t]
\centering
\includegraphics[width=0.99\columnwidth]{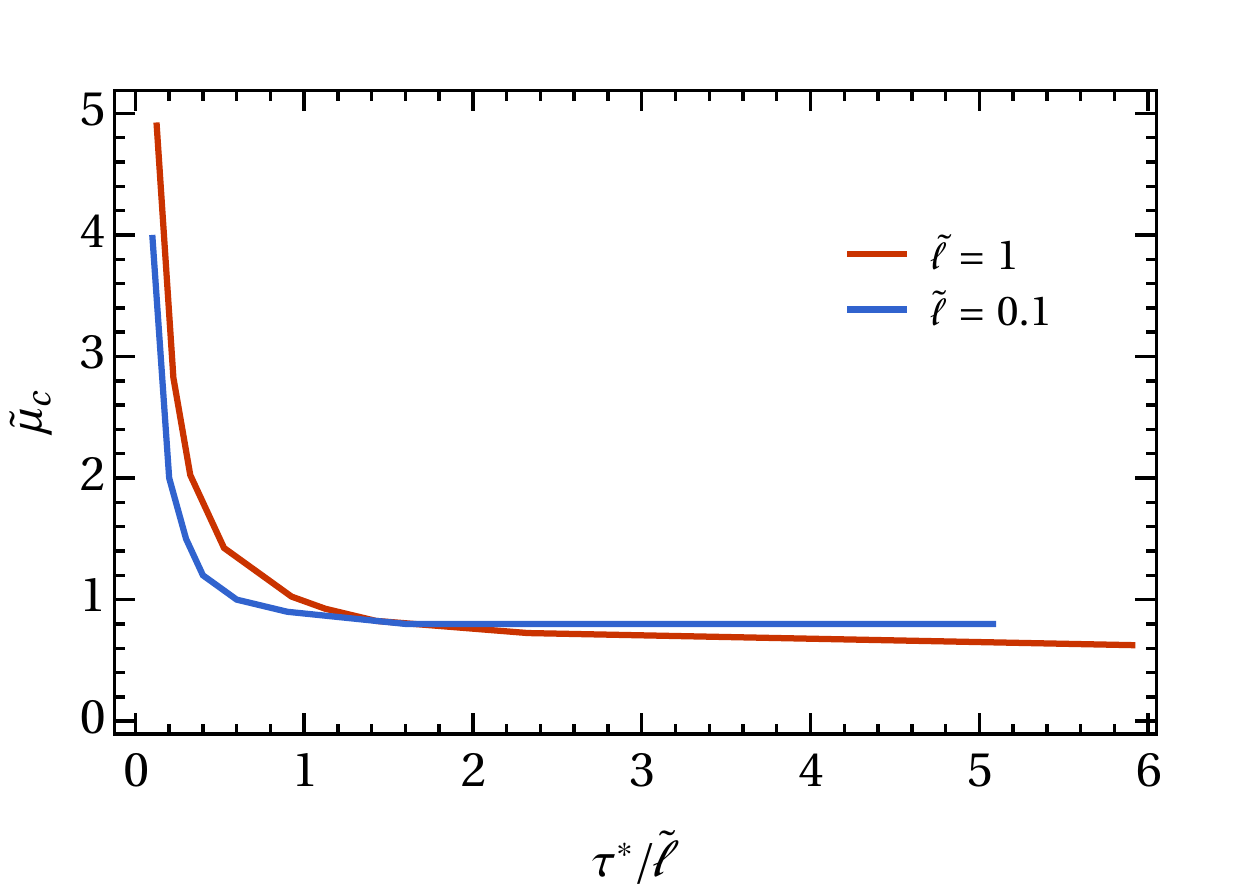}
\caption{
The phase boundary $\mut_c(\taus)$ separating successful (above curve) and failed (below) conveyor trajectories. Here $\mut = {\mu \alphat}/{\gamma D_0 L}$ is the effective mobility of the inclusion, $\lt = \ell/L$ is the length of each strip, and $\taus = D_0 \tau/L^2$ is the rescaled time interval between successive activations of strips in the accelerator; also see~\Fref{fig:accel}.
}
\label{fig:phaseDiag}
\end{figure}

\Fref{fig:phaseDiag} shows the corresponding critical mobility $\mut_c(\taus)$, i.e., the phase boundary between successful (above) and failed (below) trajectories, for $\lt = 0.1$ and $1$.  If the time interval $\taus$ between activations of successive strips in the accelerator is reduced, $\mut_c$ diverges because the particle must cover the distance $\lt$ in a shorter time-interval. On the other hand, as $\taus$ is increased, $\mut_c$ approaches a plateau because, for large $\taus$, the distance travelled by the inclusion becomes independent of $\taus$ as the density and force relax within this time interval.

\begin{figure}[t]
\centering
\includegraphics[width=0.99\columnwidth]{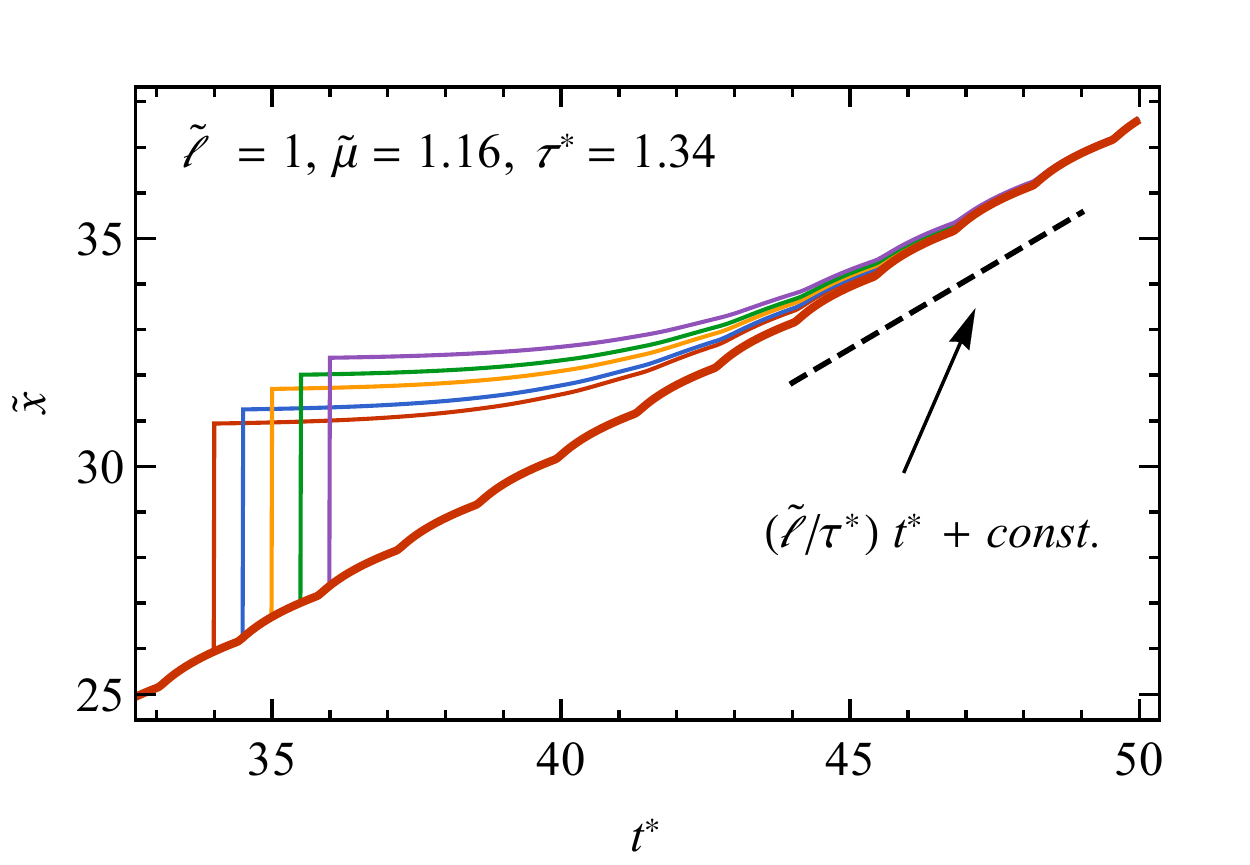}
\includegraphics[width=0.99\columnwidth]{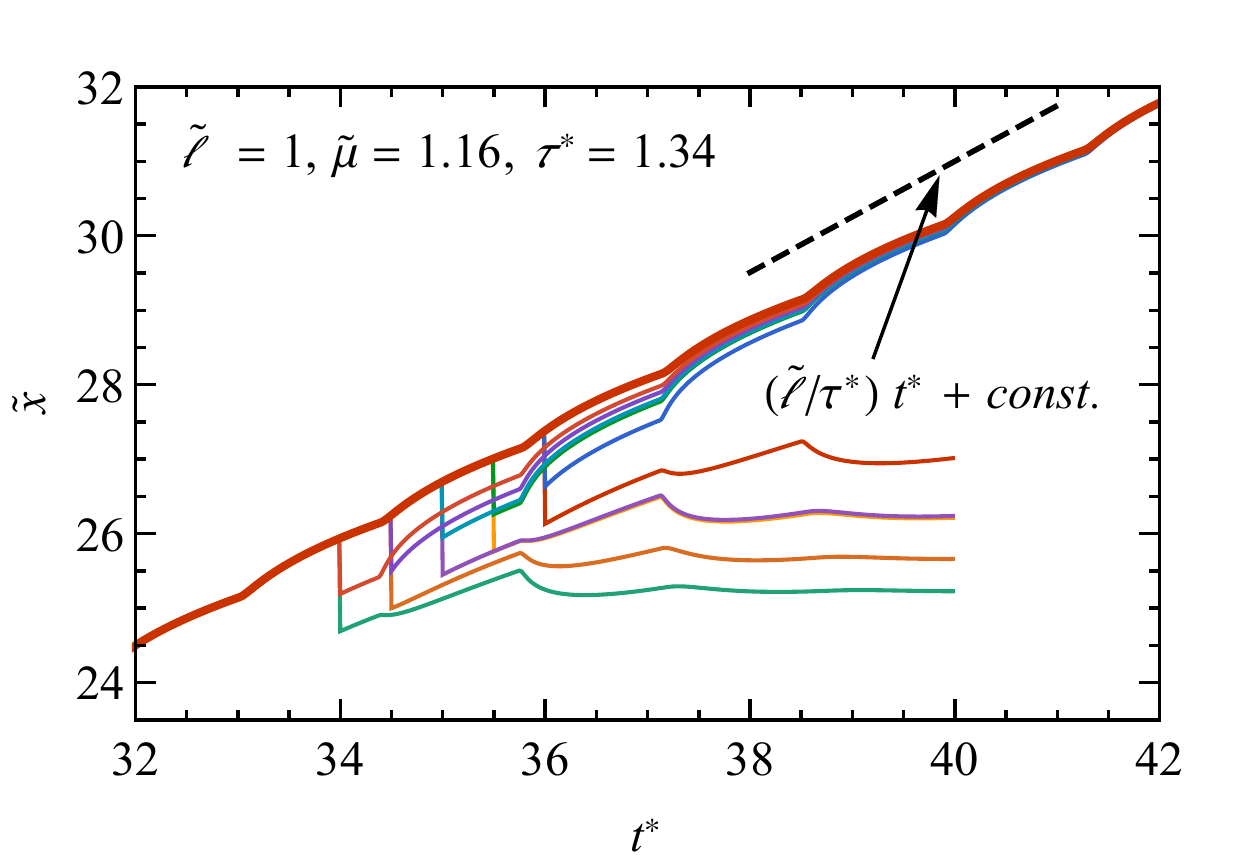}
\caption{
Abrupt perturbations of an inclusion's steady-state trajectory with velocity $\vt_{\textrm{s.s.}}=\lt/\taus$. Perturbations in the direction of motion (upper panel) always rejoin the steady state trajectory. In contrast, perturbations against the direction of motion (lower panel) may, if they are sufficiently large, expel the inclusion from the stable trajectory. Here $\xt=x/L$, $\zt=z/L=1/2$, $\ts = D_0 t/L^2$, $\lt=\ell/L = 1$, and $\taus =D_0 \tau/L^2$. }
\label{fig:trajPert}
\end{figure}

We now study the stability of trajectories  against perturbations. \Fref{fig:trajPert} shows steady-state trajectories which are perturbed by adding a finite displacement to the particle's position at some time. A steady-state trajectory is stable to arbitrarily large perturbations in the direction of motion, because the density wave simply picks up the inclusion when it catches up with the particle; see \Fref{fig:trajPert} (top). The trajectories are also stable to small perturbations opposite to the direction of motion; see \Fref{fig:trajPert} (bottom): for small perturbations, the inclusion is re-directed towards the steady-state trajectory. This, in conjunction with the aforementioned stability in the $z$-direction, confirms the expectation of stable motion on the wave's crest. For large perturbations opposite to the direction of motion, the particle ultimately comes to rest because it is overtaken by the wave front. Nonetheless, this demonstrates that a functional steady-state trajectory is a \textit{stable fixed point} of the dynamics. Clearly the range of stability vanishes as the phase boundary $\mut_c(\taus)$ in Fig.~\ref{fig:phaseDiag} is approached from above. 

We discuss in Appendix \ref{app:AppStripQuench} (also see the last paragraph of \Sref{sec:InstDensAndF}) that details of conveyor trajectories are almost independent of the $z$ component of the inclusion's position.

\subsection{Conveyance of multiple inclusions}
Imagine that each strip is deactived a time period $t_1$ after its activation. If  $t_1\gg \ell/v$, this deactivation has no effect on the motion of the inclusion which, by that time, has already traveled far away. Waiting an additional time period $t_2$ for the effect of deactivation to have ceased, where $t_2$ can be estimated from the power laws provided in \Sref{sec:pulse}, the conveyor is back at its original state. The process can thus be repeated, and    
the setup suggested in this section can be extended to convey multiple inclusions via a periodic activation protocol. This also shows that a time periodic protocol can cause net motion of particles, in contrast to the observations made at the end of \Sref{sec:pulse} above.  


\section{Discussion and outlook}
\label{sec:disc}


Spatially and temporally local perturbations (activations) lead to  long-ranged forces on inclusions in a fluid medium,
falling off as power laws with separation. Here we investigated such forces for activations that couple to diffusive modes, so that the resulting effects emerge from conservation of the fluid density. Due to wide applicability of the diffusion equation, especially on coarse-grained length scales, we expect our results and scalings to be observable in a variety of systems. Our main findings can be summarized as follows:

%

\begin{enumerate}
        \item For the geometry of two parallel, impenetrable walls at $z=0,L$ [see \Fref{fig:platepulse}], we investigated the case where an activation pulse travels along the walls with velocity $v$ over a length $\ell$. As shown in \Fref{fig:densplots}, the resulting density grows with an essential singularity at short times, and decays at late times with a power law $\sim \lt/ \sqrt{\pi \ts}$, where $\ts = D_0 t/L^2$ and $D_0$ is the diffusion coefficient of the medium.
        \item The scenario where these traveling pulses are cycled from $-\ell/2\to\ell/2\to-\ell/2$, i.e., return to their original position, is also insightful. Here we  distinguish $++$ cycles, where, during reverse motion, activation is added (i.e., the value of $\alpha$ is the same for both parts of the cycle), and $+-$ pulses, which have a sign change of $\alpha$ so that the  reverse pulse reverts the system to its original state. The resulting density in \Fref{fig:densplotscycle} shows strong dependencies on the $x$-coordinate, and generally decays more rapidly for a $+-$ cycle, since leading contributions cancel in this case. The associated force, shown in \Fref{fig:forcecycle}, also exhibits subtle position dependence, decaying most rapidly $\sim (\ts)^{-7/2}$ at the geometric midpoint for a $++$ pulse. Away from this point, $+-$ pulses decay $\sim (\ts)^{-5/2}$ while $++$ pulses decay $\sim (\ts)^{-3/2}$ at late times. Exact analytical forms for these late-time asymptotes [see \Eref{eq:FcycleStatAsymp}] highlight the underlying cancellations.
        \item The net force experienced by a stationary inclusion during a $+-$ pulse (i.e., if no net density is injected into the system) is zero, reminiscent of the scallop theorem for swimming at low Reynolds numbers~\cite{purcell1977LowRe,najafi2005PropLowRe,lauga2009HydrodynSwim}.
        \item We investigated the work $W$ that can be extracted from the motion of an inclusion on a closed trajectory between the two walls, subject to a cyclic pulse activation of the form of point 2 above. $W$ can be positive or negative, depending on the dimensionless frequency $\omegat = \omega L^2/D_0$ of the inclusion's motion and dimensionless speed $\vt={v}L/{D_0}$ of the pulse. For rapid pulses, reversing the trajectory changes the sign of $W$, while for slower pulses, an intricate dependence on $\omegat$ is observed; see \Fref{fig:WorkClosedTraj}. 
        \item The functionality of a \textit{conveyor}, as illustrated in \Fref{fig:accel}, was investigated, using a special version of the above-mentioned activation protocol for parallel walls: regions of length $\ell$ are activated successively, a time interval $\tau$ apart. This gives rise to a density wave traveling at an average speed of $v=\ell/\tau$, which is able to propel small inclusions in its wake. The associated trajectories (see Figs.~\ref{fig:trajectories} and \ref{fig:trajPert}) were shown to be stable against perturbations, so that the conveyor is functional for sufficiently large (dimensionless) values of the inclusion's mobility $\mut={\mu \alphat}/{\gamma D_0 L}$ ($\gamma$ is its friction coefficient).
        \item The existence of stable trajectories is reliant on time-dependence of the  density. This distinguishes the conveyor mechanism from diffusiophoretic forces in steady state.
        \item A critical mobility $\mut_c(\tau)$ separates functional from non-functional trajectories of the conveyor; see \Fref{fig:phaseDiag}. 
\end{enumerate}

As has been demonstrated via simulations of active Brownian particles, quenches of the \textit{activity} in an active fluid give rise to effects that are captured qualitatively by the density-based models described here. Therefore we expect our results to be relevant and of interest for active matter systems~\cite{solon_active_2015, berthier2015epl,catesX}, but also suspensions, where interactions can be tuned~\cite{maretkeim2004,ballauff2006}.

In future work we plan to study the above effects in other geometries, and to explore fluctuation-induced effects related to time-dependent quenching protocols. It would further be interesting to investigate these effects for activations which couple to additional conserved modes (e.g., momentum).

\begin{acknowledgments}
CMR gratefully acknowledges Prof.~S.~Dietrich for support. M.~Kardar is supported by the NSF through Grant No. DMR-1708280. 
\end{acknowledgments}

\pagebreak

\onecolumngrid

\appendix
\section{Detailed study of the density and forces arising from an activated strip pair}
\label{app:AppStripQuench}

Here we discuss the density and forces as a function of position (for various times) between two walls after a single strip in each wall is activated. The walls are situated at $\tilde z=z/L=0,1$, and each wall has a single strip of length $\lt=\ell/L=1$ in the region $\xt=x/L \in [-0.5,0.5]$ with $\tilde z=z/L=0,1$, respectively.

\begin{figure}[h]
\centering
\includegraphics[width=0.195\textwidth]{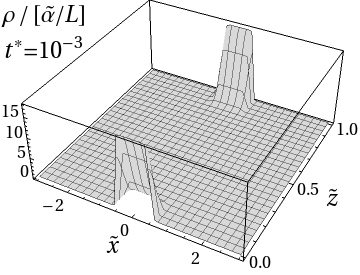}
\includegraphics[width=0.195\textwidth]{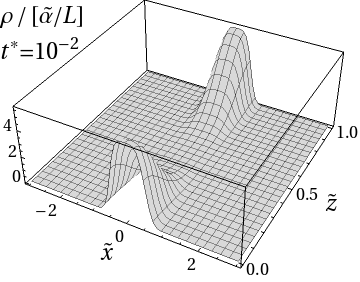}
\includegraphics[width=0.195\textwidth]{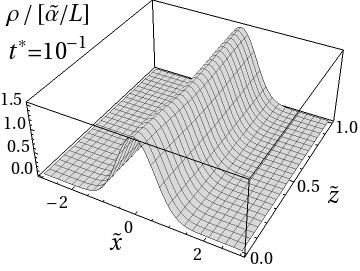}
\includegraphics[width=0.195\textwidth]{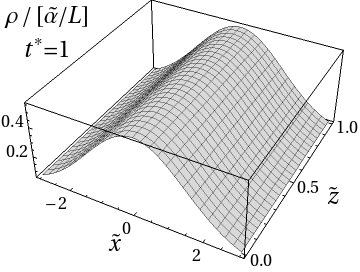}
\includegraphics[width=0.195\textwidth]{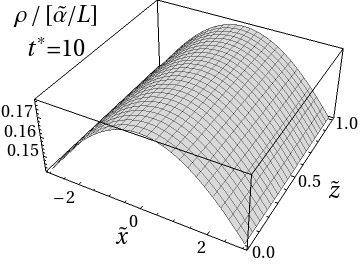}
\hspace{0.33\textwidth}
\caption{Density $\rho$ between two walls, separated by a distance $L$ along the $z$ axis, at different times $\ts = D_0t/L^2 $ after activating the strips. Each wall has a single strip of length $\lt=\ell/L=1$ in the region $\xt=x/L \in [-0.5,0.5]$ with $\tilde z=z/L=0,1$, respectively.
}
\label{fig:rho}
\end{figure}

\Fref{fig:rho} shows the density $\rho$ at different times. At early times, $\rho$ is peaked near the activated strips; these peaks then diffuse into the system. For $\ts = D_0t/L^2\gtrsim10^{-1} $, the density is essentially gradient-free in the transverse $z$ direction, and further relaxation occurs only along the lateral $x$ direction.

\begin{figure}[h]
\centering
\includegraphics[width=0.195\textwidth]{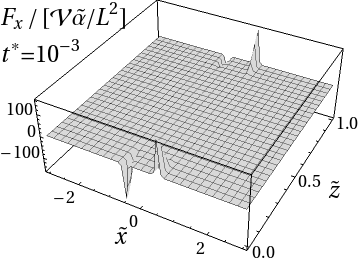}
\includegraphics[width=0.195\textwidth]{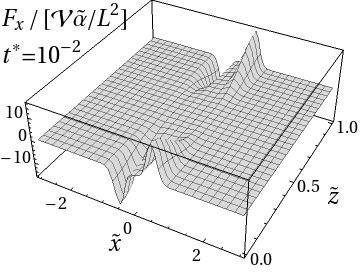}
\includegraphics[width=0.195\textwidth]{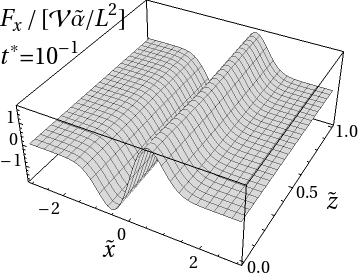}
\includegraphics[width=0.195\textwidth]{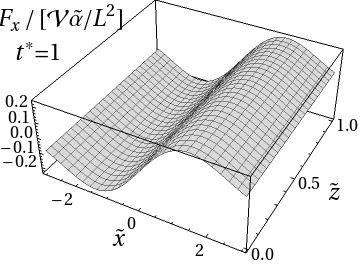}
\includegraphics[width=0.195\textwidth]{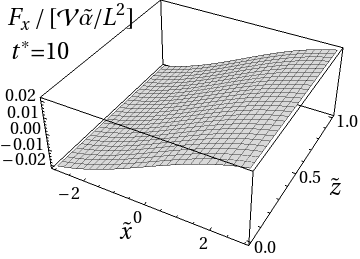}
\hspace{0.33\textwidth}
\caption{Lateral component $F_x$ of the force on an inclusion of volume $\mathcal V$, situated between two walls that are separated by a distance $L$ along the $z$ axis, at different times $\ts = D_0t/L^2 $ after activating the strips. Each wall has a single strip of length $\lt=\ell/L=1$ in the region $\xt=x/L \in [-0.5,0.5]$ with $\tilde z=z/L=0,1$, respectively.
}
\label{fig:Fx}
\end{figure}

\Fref{fig:Fx} shows the lateral component $F_x$ that would be experienced by an inclusion, as a function of position, at different times. Naturally this force is anti-symmetric about the line $x=0$. As with the density, $F_x$ is peaked near the activated strips at early times. As soon as the density has relaxed along the transverse $z$ direction ($\ts = D_0t/L^2\gtrsim10^{-1} $), $F_x$ becomes independent of $z$.

\begin{figure}[h]
\centering
\includegraphics[width=0.195\textwidth]{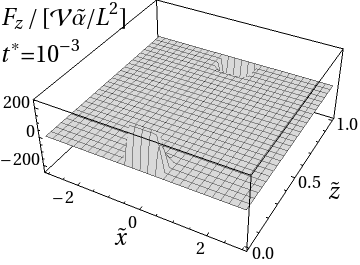}
\includegraphics[width=0.195\textwidth]{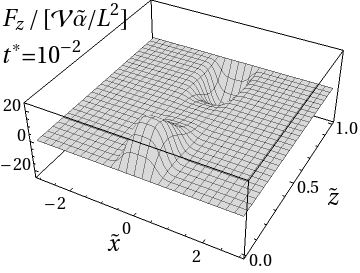}
\includegraphics[width=0.195\textwidth]{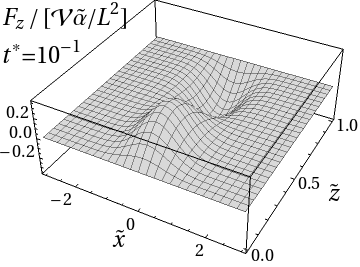}
\includegraphics[width=0.23\textwidth]{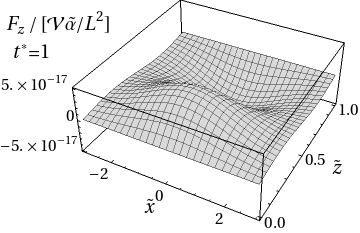}
\hspace{0.33\textwidth}
\caption{Perpendicular component $F_z$ of the force on an inclusion of volume $\mathcal V$, situated between two walls that are separated by a distance $L$ along the $z$ axis, at different times $\ts = D_0t/L^2 $ after activating the strips. Each wall has a single strip of length $\lt=\ell/L=1$ in the region $\xt=x/L \in [-0.5,0.5]$ with $\tilde z=z/L=0,1$, respectively.
}
\label{fig:Fz}
\end{figure}

Finally, \Fref{fig:Fz} shows the transverse component $F_z$ that would be experienced by an inclusion, as a function of position, at different times. By symmetry, for points along the central line $\zt=1/2$, $F_z$ always points towards the central line. Moreover, for $\ts \gtrsim10^{-1} $, $F_z$ decays very rapidly to zero, because the density has relaxed along the $z$ direction.

From this analysis we conclude that an inclusion situated along the central line $\zt=1/2$, will maintain this $z$ coordinate for all times. For points sufficiently close to $\zt=1/2$, $F_z$ essentially acts as a trap force that restores the inclusion towards the central line. It is for this reason that the trajectories of inclusions in the accelerator (see \Sref{sec:accel}) are treated as one-dimensional along the line $\zt=1/2$. \\

\begin{figure}[t]
\centering
\includegraphics[width=0.455\textwidth]{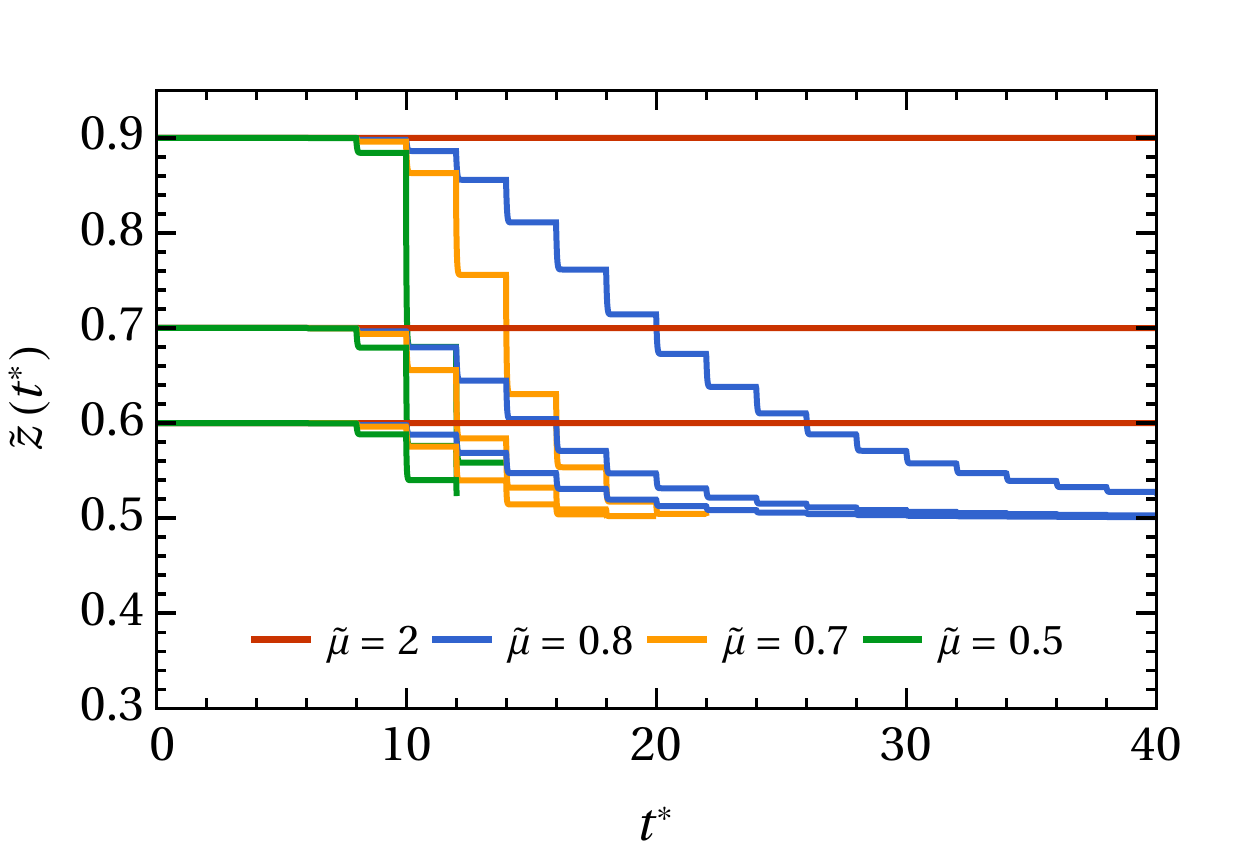}
\caption{
Transverse ($z$) component of the full $2D$ trajectory of an inclusion which is impinged upon by a steady-state wave-front from the left, for various mobilities  $\mut={\mu \alphat}/{\gamma D_0 L}$, and various initial $\zt$. Here $\zt=z/L$, $\ts = D_0 t/L^2$, $\lt = 1$, and $\taus = 2$ (as used in \Fref{fig:trajectories}). 
}
\label{fig:ztraj}
\end{figure}

In order to check this explicitly, we computed full 2D trajectories for various initial coordinates away from $\zt=0.5$. The results are shown in \Fref{fig:ztraj}, for the same parameters as were used in \Fref{fig:trajectories}, for which $\mut_c\simeq0.75$. We observe that indeed $F_z$ tends to push the particle towards $\zt=1/2$ when $\mut\simeq\mut_c$. When $\mut\gg\mut_c$, the inclusion remains so far ahead of the density wave (in the $x$ direction), that it feels a vanishing $F_z$, and thus its $\zt$ coordinate remains unchanged. We have also checked that the $\xt$ trajectories obtained for these cases are mutually indistinguishable, and jointly indistinguishable from those obtained for \Fref{fig:trajectories} where the particle began at $\zt=1/2$. 

This justifies treating the accelerator of \Sref{sec:accel} as an essentially one dimensional problem, where only the $\xt$ trajectories are computed.

\pagebreak

\twocolumngrid

\bibliographystyle{apsrev4-1}   
\bibliography{references}

\end{document}